\documentclass[longauth]{aa} 

\usepackage{graphicx}
\usepackage{txfonts}
\usepackage[version=4]{mhchem}
\usepackage{upgreek}
\usepackage{chemformula}
\usepackage{multirow}
\begin{document} 

   \title{Modelling methanol and hydride formation in the JWST Ice Age era}


\titlerunning{Modelling CH$_3$OH and hydride formation in the JWST Ice Age era}
\authorrunning{I. Jim\'enez-Serra et al.}

   \author{Izaskun Jim\'enez-Serra\inst{1}, 
           Andrés Megías\inst{1},
           Joseph Salaris\inst{2},
           Herma Cuppen\inst{2},
           Angèle Taillard\inst{1},
           Miwha Jin\inst{3,4},
           Valentine Wakelam\inst{5},
           Anton I. Vasyunin\inst{6},
           Paola Caselli\inst{7},
           Yvonne J. Pendleton\inst{8},
           Emmanuel Dartois\inst{9},
           Jennifer A.~Noble\inst{10},
           Serena Viti\inst{11,12,13},
           Katerina Borshcheva\inst{6,14},
           Robin T. Garrod\inst{15},
           Thanja Lamberts\inst{16,11},
          Helen Fraser\inst{17}, Gary Melnick\inst{18},
          Melissa McClure\inst{11},
          Will Rocha\inst{11},
          Maria N. Drozdovskaya\inst{19} 
          \and
          Dariusz C.  Lis\inst{20}
          }
   \institute{Centro de Astrobiolog\'{\i}a (CAB), CSIC-INTA, Ctra. de Ajalvir km 4, E-28850, Torrej\'on de Ardoz, Spain\\
              \email{ijimenez@cab.inta-csic.es} 
         \and
         Institute for Molecules and Materials, Radboud University, Heyendaalseweg 135, 6525 AJ, Nijmegen, The Netherlands
         \and
         Astrochemistry Laboratory, Code 691, NASA Goddard Space Flight Center, Greenbelt, MD 20771, USA
         \and
         Department of Physics, Catholic University of America, Washington, DC 20064, USA
         \and
         Laboratoire d’astrophysique de Bordeaux, Univ.Bordeaux, CNRS, B18N, allée Geoffroy Saint-Hilaire, 33615 Pessac, France
         \and
         Institute of Natural Sciences and Mathematics, Ural Federal University, 19 Mira Str., 620075 Ekaterinburg, Russia
         \and
         Max-Planck-Institut f\"ur extraterrestrische Physik, Gießenbachstrasse 1, D-85748 Garching bei M\"unchen, Germany
         \and
         Department of Physics, University of Central Florida, Orlando, FL 32816, USA
         \and
         Institut des Sciences Mol\'eculaires d’Orsay, CNRS, Univ. Paris-Saclay, 91405 Orsay, France
         \and
         Physique des Interactions Ioniques et Mol\'{e}culaires, CNRS, Aix Marseille Univ., 13397 Marseille, France
         \and
         Leiden Observatory, Leiden University, PO Box 9513, NL-2300 RA Leiden, the Netherlands
         \and 
    Transdisciplinary Research Area (TRA) 'Matter'/Argelander-Institut f\"ur Astronomie, University of Bonn, Germany
        \and
        Physics and Astronomy, University College London, Gower Street, London WC1E 6BT, UK
        \and
         Institute of Astronomy of the RAS, Pyatnitskaya st. 48, 119017, Moscow
         \and
         Department of Astronomy and Chemistry, University of Virginia, Charlottesville, VA, 22904, USA
         \and
         Leiden Institute of Chemistry, Gorlaeus Laboratories, Leiden University, P.O. Box 9502, 2300 RA Leiden, The Netherlands
        \and 
        School of Physical Sciences, The Open University, Kents Hill, Milton Keynes MK7 6AA, UK. 
        \and
        Center for Astrophysics | Harvard \& Smithsonian, 160 Concord Avenue, Cambridge 02138 MA, USA.
        \and
        Physikalish-Meteorologisches Observatorium Davos und Weltstrahlungszentrum (PMOD/WRC), Dorfstrasse 33, CH-7260, Davos Dorf, Switzerland
        \and 
        Jet Propulsion Laboratory, California Institute of Technology, 4800 Oak Grove Drive, Pasadena, CA 91109, USA}

   \date{Received September 15, 1996; accepted March 16, 1997}

 
  \abstract
   {Recent JWST observations have measured the ice chemical composition toward two highly-extinguished background stars, NIR38 and J110621, in the Chamaeleon I molecular cloud. The observed excess of extinction on the long-wavelength side of the H$_2$O ice band at 3$\,$$\mu$m has been attributed to a mixture of CH$_3$OH with ammonia hydrates (NH$_3\cdot$H$_2$O), which suggests that CH$_3$OH ice in this cloud could have formed in a water-rich environment with little CO depletion. 
   Laboratory experiments and quantum chemical calculations suggest that CH$_3$OH could form via the grain surface reactions CH$_3$ + OH and/or C + H$_2$O in water-rich ices. However, no dedicated chemical modelling has been carried out thus far to test their efficiency. In addition, it remains unexplored how the efficiencies of the proposed mechanisms depend on the astrochemical code employed.}
   {We model the ice chemistry in the Chamaeleon I cloud to establish the dominant formation processes of CH$_3$OH, CO, CO$_2$, and of the hydrides CH$_4$ and NH$_3$ (besides H$_2$O). By using a set of state-of-the-art astrochemical codes (MAGICKAL, MONACO, Nautilus, \textsc{Uclchem}, and KMC simulations), we can test the effects of the different code architectures (rate equation vs. stochastic codes) and of the assumed ice chemistry (diffusive vs. non-diffusive).}
   {We consider a grid of models with different gas densities, dust temperatures, visual extinctions, and cloud-collapse length scales. Besides the successive hydrogenation of CO, the codes chemical networks have been augmented to include the alternative processes for CH$_3$OH ice formation in water-rich environments (i.e. the reactions CH$_3$ + OH $\rightarrow$ CH$_3$OH and C + H$_2$O $\rightarrow$ H$_2$CO).}
   {Our models show that the JWST ice observations are better reproduced for gas densities $\geq$10$^5$ cm$^{-3}$ and collapse time-scales $\geq$10$^5$ yr. CH$_3$OH ice formation occurs predominantly ($>$99\%) via CO hydrogenation. The contribution of reactions CH$_3$ + OH and C + H$_2$O, is negligible. The CO$_2$ ice may form either via CO + OH or CO + O depending on the code. However, KMC simulations reveal that both mechanisms are efficient despite the low rate constant of the CO + O surface reaction. CH$_4$ is largely underproduced for all codes except for \textsc{Uclchem}, for which a higher amount of atomic C is available during the translucent cloud phase of the models. Large differences in the predicted abundances are found at very low dust temperatures (T$_{\rm dust}$$<$12 K) between diffusive and non-diffusive chemistry codes. This is due to the fact that non-diffusive chemistry takes over diffusive chemistry at such low T$_{\rm dust}$. This could explain the rather constant ice chemical composition found in Chamaeleon I and other dense cores despite the different visual extinctions probed.}
  {}


   \maketitle
%

\section{Introduction}
\label{sec:intro}

Interstellar ices represent the main reservoir of organic and volatile components in molecular clouds and star-forming regions \citep[see e.g. the review by][and references therein]{Boogert2015}. Ices grow during the collapse of molecular clouds, largely influencing the chemical composition of protoplanetary disks and, ultimately, of planetesimals and cometesimals \citep[see e.g.][]{Quenard2018b, Coutens2020}. Indeed, a fraction of the chemical compounds measured in comet 67P/Churyumov-Gerasimenko by the ESA {\it Rosetta} mission could be of prestellar origin, which suggests that they could have formed during the dense molecular cloud phase \citep{Altwegg2016, Drozdovskaya2019, Drozdovskaya2021}. 

The composition of ices in molecular clouds and star-forming regions is studied through absorption spectroscopy of interstellar ices in the near- and mid-IR against bright background sources \citep[see e.g.][] {Whittet1983,Pendleton1999,Boogert2013,Goto2018}. However, these observations have been scarce because they are challenging from the ground due to telluric features and high sky-background noise in the IR. In addition, dense molecular clouds and dense cloud cores can have very high levels of visual extinction ($A_{\rm V}$ $\geq$ 20 mag), for which detailed ice absorption spectroscopic observations in the near- and mid-IR require IR-bright background field stars and very high sensitivity. 

From the ground, H$_2$O, CO and methanol (CH$_3$OH) ices have been detected toward a small sample of dense molecular clouds such as Taurus, Lupus, Serpens and the Pipe Nebula \citep{Whittet1983,Boogert2013,Goto2018,Perotti2020} and toward dense cloud cores such as the star-forming core L483, and the L1544 and L694-2 prestellar cores \citep{Chu2020,Goto2021}. From space, the Infrared Space Observatory ({\it ISO}), the {\it Spitzer Space Telescope}, and the AKARI Infrared Satellite, provided the first look at the chemical composition of ices in the mid-IR in dense molecular clouds and cores \citep{Whittet1998,Dartois2005,Knez2005,Whittet2009,Boogert2011,Noble2013}, which included information on additional ysosmolecular species such as CO$_2$ or OCN$^-$.  
However, the sensitivity or  spectral resolving power of these observations was rather limited, 
which allowed only the detection of simple and abundant species toward lines-of-sight in molecular clouds with moderate levels of visual extinction  \citep[$A_{\rm V}$ $\leq$ 30 mag; see e.g. the summary by][]{Boogert2015}. 

The {\it James Webb Space Telescope} (JWST), with its unprecedented sensitivity and high spectral resolving power (up to $\sim$3000), is providing the most comprehensive view of the chemical composition of interstellar ices during the initial conditions of star and planet formation. The Ice Age Director's Discretionary Early Release Science program has obtained ice absorption spectra toward hundreds of field stars located behind the Chameleon I dense molecular cloud \citep{McClure2023}. The goal of the project is to establish the chemical composition of interstellar ices concurrent to the process of Solar-type system formation from the molecular cloud stage, to prestellar cores, young stellar objects (Class 0 and I sources), and, ultimately, protoplanetary disks \citep[Class II-type sources;][]{McClure2023, sturm2023,sturm2024,dartois2024,Noble2024}. The first results have reported the ice chemical composition toward the most highly-extinguished background stars ever observed to date in molecular clouds, NIR38 and J110621, with pre-flight estimated visual extinctions of $A_{\rm V}$ = 60 mag and $A_{\rm V}$ = 95 mag, respectively \citep[][]{McClure2023}, although these values are subject to large uncertainties \citep[see][and Section$\,$\ref{sec:Av}]{dartois2024,Noble2024}. 
These observations reveal that, the ice composition is very similar toward the two lines-of-sight: CO ice is $\sim$30\% and $\sim$45\% by number, respectively, relative to water; and CO is more abundant than CO$_2$ (between 13\% and 20\% with respect to water) and methanol (between 4 and 9\%). Hydrides such as CH$_4$ and NH$_3$ comprise only up to 5\% of ice with respect to water. 

Based on the expected total CO abundance in molecular clouds for these levels of extinction \citep{lacy2017}, it was proposed that the carbon-bearing ices observed with the JWST accounted for $\leq$50\% of the total molecular CO in gas or ice \citep{McClure2023}.
This scenario was consistent with the prominent absorption feature in the red wing of the water 3$\mu$m band, which was attributed to the presence of hydrides suggesting a water-rich ice environment \citep[see Extended Data Figure$\,$9 in][]{McClure2023}. Under these conditions, the formation of methanol at 10-20 K is suggested to be driven by the grain surface reaction chain CH$_4$ + OH $\rightarrow$ CH$_3$ + H$_2$O and CH$_3$ + OH $\rightarrow$ CH$_3$OH, as proposed by laboratory experiments and quantum chemical calculations of ice formation \citep{Qasim18,molpeceres2021carbon}. 
Additionally, the \ce{CO2} ice feature profile was better matched with laboratory spectra of a \ce{CO2}-\ce{H2O} ice blend \citep[Extended Data Figure$\,$4; see][]{McClure2023}, and weak features between 7-7.5$\,$$\mu$m were consistent with ethanol blended with \ce{H2O} suggesting a similar formation environment for these species. The formation of hydrides such as CH$_4$ and NH$_3$ could also proceed in water-rich ice environments under the same cold interstellar conditions as shown by laboratory experiments \citep[][]{Hiraoka1995,Qasim2020,Lamberts2022}, which supports the scenario of efficient CH$_3$OH ice formation in a water-ice rich environment toward the lines of sight of NIR38 and J110621. 

This idea, however, has been re-evaluated recently and new radiative transfer calculations show that much of the red wing of the 3 $\mu$m water band observed with JWST toward these two field stars, could alternatively be due to grain growth and not hydrides, suggesting that ices in the Chamaeleon I cloud may not be as water-rich as initially thought \citep{dartois2024}. In that case, the production of methanol via the reaction CH$_3$ + OH would be less efficient, and thus other chemical mechanisms for methanol formation \citep[as e.g. the successive hydrogenation of CO; see][]{Watanabe2002} are expected to take over. It remains unclear, however, in what amount these two mechanisms contribute to the production of methanol in ices and hence, dedicated chemical modelling is needed to investigate the possible and relative importance of different chemical routes for the production of methanol ice.


In a recent work, \citet{jin2022ice} modelled the ice chemistry toward the Class 0 source Cha-MMS1 also observed within the Ice Age program, where the gas densities reach values $\geq$10$^{10}$$\,$cm$^{-3}$ and gas/dust temperatures are $\geq$100$\,$K. However, this modelling did not focus on the formation of ices toward the quiescent NIR38 and J110621 sightlines in the Chamaeleon I molecular cloud, with typical gas densities of $\sim$10$^4$-10$^6$$\,$cm$^{-3}$ and temperatures of $\sim$10-15$\,$K, and therefore, the chemical modelling of the JWST data toward these two background stars is still missing.
In addition, while different chemical codes do exist at present, they are built following different approaches and techniques (rate-equation approach vs. stochastic Monte Carlo techniques) or they consider different chemical processes on the surface of dust grains \citep[diffusive vs. non-diffusive reaction mechanisms and thermal and non-thermal desorption processes; see e.g. the review by][]{Cuppen2017}. The Ice Age results therefore provide a unique opportunity to compare different astrochemical codes so that the dominant chemical routes yielding the ice chemical inventory detected with JWST toward the Chamaeleon I molecular cloud, can be constrained.  

In this work, we model the chemical composition of the ices observed with the JWST toward the background stars NIR38 and J110621 within the Ice Age ERS program. For this modelling effort, we use five different astrochemical codes to identify the features (or chemical processes)  within these models that best reproduce the observed ice chemical compositions. Note that the last community effort to compare astrochemical models goes back to the review work of \citet{Cuppen2017}. 

The paper is organised as follows. In Section \ref{sec:models} we describe the astrochemical models used and the parameter space of the models used in this work. In Section \ref{sec:results}, we report the results of the models and compare the ice molecular abundances predicted by the models with those measured with the JWST. In Section \ref{sec:best}, we identify the models from each astrochemical code that best reproduce the ice chemical compositions observed within Ice Age. Section \ref{sec:lowTdust} discusses how our results change if lower dust temperatures are considered in our best models. In Section \ref{sec:discussion} we present the general trends obtained from the models and discuss the uncertainties and caveats. Finally, our conclusions are summarised in Section \ref{sec:conclusions}.

\section{Description of the chemical models} \label{sec:models}

Our aim is to model the ice chemical formation and evolution in the Chameleon I dense molecular cloud using different astrochemical codes. These codes are built following different philosophies: we have both rate equation codes such as MAGICKAL \citep{garrod19}, MONACO \citep{VasyuninHerbst13,Vasyunin_ea17}, Nautilus \citep{Ruaud2016} and \textsc{Uclchem} \citep{Holdship2017}; as well as the stochastic Kinetic Monte Carlo (KMC) model of \citet{Cuppen2007}. These codes also include different levels of complexity for the formation and chemistry of the ices and, thus, the idea of this work is to test the features from these astrochemical codes that best reproduce the JWST observations obtained toward NIR38 and J110621, which are field stars with the highest extinction observed thus far \citep{McClure2023}. For instance, while MAGICKAL, MONACO and KMC include diffusive and non-diffusive chemistry, Nautilus and \textsc{Uclchem} only include diffusive chemistry. On the other hand, Nautilus and \textsc{Uclchem} include some non-thermal desorption mechanisms that are not considered in the other codes. See Table$\,$\ref{table-comparison-codes} for the main characteristics and features of each of these codes. In this section, we briefly describe the astrochemical codes employed in this work (listed in alphabetical order) and we report the physical parameter space used to run the models. 

\begin{table*}
\setlength{\tabcolsep}{10pt} 
\caption{Main features of the astrochemical codes used in this work. \label{table-comparison-codes}} 
\centering           
\begin{tabular}{lccccc}  
\hline   
& {\bf MAGICKAL} & {\bf MONACO} & {\bf Nautilus} & {\bf \textsc{Uclchem}} & {\bf KMC} \\ \hline
{\bf Code type} & Rate Equation & Rate Equation & Rate Equation & Rate Equation & Stochastic \\ \hline 
{\bf Thermal desorption} &  \checkmark & \checkmark & \checkmark & \checkmark & \checkmark \\ \hline
{\bf Non-thermal desorption} & & & & & \\
UV photodesorption & \checkmark & \checkmark & \checkmark & \checkmark & \checkmark \\
CR-induced photodesorption & \checkmark & \checkmark & \checkmark & \checkmark & \checkmark \\
CR-induced heating & \checkmark & \checkmark & \checkmark & \checkmark & \checkmark \\
Chemical (reactive) desorption & \checkmark & \checkmark & \checkmark & \checkmark & \checkmark \\ 
H$_2$-formation-induced desorption & $-$ & $-$ & $-$ & \checkmark & $-$ \\
CR sputtering & $-$ & $-$ & \checkmark & $-$ & $-$ \\ \hline
{\bf Diffusive chemistry} &  \checkmark & \checkmark & \checkmark & \checkmark & \checkmark \\ \hline
{\bf Non-diffusive chemistry} &  \checkmark & \checkmark & $-$ & $-$ & \checkmark \\ \hline
\end{tabular}
\end{table*}

\subsection{Kinetic Monte Carlo (KMC) simulations}
\label{sec:kinetic}
The KMC method denotes a class of algorithms aimed at modelling the time evolution of a given system by generating a stochastic trajectory using random sampling. This is the only code of the stochastic type used in this work. This method provides detailed information about the microscopic processes taking place on the surface of dust grains. For an overview of the theoretical background and its application to interstellar ice layers, see \citet{Cuppen2013}. 

In the present study, a microscopic type of KMC was used called "continuous-time random-walk kinetic Monte Carlo" (CTRW-KMC). This particular simulation program is an adaptation from the one used in \citet{Cuppen2007}, \citet{Cuppen2009}, \citet{Lamberts2014},  and \citet{Simons2020}. The program exclusively simulates grain surface chemistry, but the results can be coupled to the gas by submitting gas-phase abundances as input, from which the adsorption fluxes are calculated. For the present paper, the code was adapted to be able to treat densities and gas abundances that vary with time. These were simulated using the \textsc{Uclchem} code, where collapse was assumed to follow free-fall and the input parameters were varied in accordance with the physical parameter grid of Table \ref{tab:gridofmodels} and described in Section \ref{sec:parameters}. The resulting time-dependent densities and a selection of gas abundances were then passed on to the KMC grain code. Note that there is no feedback from the thermal desorption of grain species onto the gas abundances, as we assume the effect to be negligible at these low temperatures. The selection of gas abundances transferred to the KMC model was limited to a small set of species: H, CO, O, C and N, i.e. the atomic and molecular species that contribute to the formation of the main ice species studied here (CO$_2$, CH$_3$OH, and the hydrides). The selection was kept minimal, because we do not expect the results from the \textsc{Uclchem} grain surface chemistry to conflate with those from the KMC simulations. This is due to the low dust temperatures and UV radiation fields considered in our models, which induce little desorption from the grain surface.

The surface of the grain is represented by a $50\times 50$ lattice with periodic boundary conditions upon which species can accrete on top and layers can form. Apart from thermally activated processes such as diffusion (hopping) and desorption, the lattice KMC model also includes the non-thermal processes of direct photodissociation and cosmic-ray dissociation. Diffusion and desorption are made site-specific processes where the binding energy is neighbour-dependent. The sequence of processes and corresponding time advances are determined in a stochastic manner from the rates. See \citet{Cuppen2007} for a more detailed account of the calculation of the rate coefficients for these various processes. The chemical reaction networks used were confirmed experimentally or by quantum mechanical calculations in \citet{Lamberts2014} (water network), \citet{Simons2020} (CO network) and \citet{Ioppolo2020} (glycine network).

\subsection{MAGICKAL code}
\label{sec:magickal}

The MAGICKAL code is a three-phase astrochemical model that solves a set of rate equations describing chemical kinetics in the gas, on the grain surface, and within the bulk of the ice mantle. Grain-surface chemistry is coupled with the gas through accretion and desorption of chemical species. The latter can proceed both thermally and non-thermally. Non-thermal desorption mechanisms include chemical (reactive) desorption, and photo-desorption by both external and CR-induced UV photons.

In this model, grain-surface chemistry is controlled by both diffusive and non-diffusive mechanisms. The former represents the chemical kinetics described by the thermal diffusion of surface species by which surface species may meet and react. MAGICKAL uses the modified-rate method of \citet{garrod2008b} for diffusive surface reactions \citep[see also][]{garrod2009}; this method adjusts reaction rates to replicate the stochastic behaviour of the system when particular reactions fall into the so-called “small-grain” limit, i.e. when reactant populations are small and the diffusion rate is large. In these cases, the reaction rate should be limited by the accretion (or production) rate of the reactants on the surface.

The diffusion rate is governed by the temperature of the dust and the diffusion barrier ($E_\textrm{dif}$) of the relevant species. The diffusion barriers of the surface species are assumed to be some constant fraction of their binding energies to the surface ($E_\textrm{des}$), initially based on work on metal surfaces where often a ratio between the diffusion barrier and the binding energy of 0.3 was observed. However, astrochemically relevant systems appear to show a larger scatter for these data based on the few systems for which both the diffusion barrier and the binding energy are known \citep{Karssemeijer2014,Yang2022}. Because of this, astrochemical models usually assume a fixed value in the approximate range of 0.3-0.8. In MAGICKAL, we assumed 0.35 for the molecules, and 0.55 for other atomic species, as these values best reproduce the production of interstellar \ce{CO2} ice~\citep{garrodandpauly11}. For atomic H, average binding energy and diffusion barrier values of 661$\,$K and 243$\,$K, respectively, are used, based on the calculations of \citet{Senevirathne2017}. The bulk diffusion barriers for H and H$_2$ are twice the values used for surface diffusion, as per \citet{garrod2013}.

The kinetics of non-diffusive mechanisms consider the probability by which a newly formed chemical species reacts with other potential reactants in the proximity without diffusion. This mechanism is important for accurately treating the production of complex organic molecules (COMs) at low temperatures~\citep{garrod19,jinandgarrod20}. For the details of the four non-diffusive mechanisms included in MAGICKAL, we refer to \citet{garrod22}.

The surface layer is coupled to the bulk ice phase through the net gain/loss rate of material to/from the surface, which either acts to cover up the existing surface or to expose the underlying bulk ice material \citep{hasegawaandherbst93, garrodandpauly11}. The chemical kinetics within the bulk ice is controlled by both tunnelling-mediated bulk diffusion for H and H$_2$~\citep[see][]{garrod22} and non-diffusive chemistry. The chemical networks used in these models are based on that presented by \citet{garrod22}.  

\subsection{MONACO code}
\label{sec:monaco}
The gas-grain astrochemical model MONACO \citep[][Borscheva et al., in prep.]{Vasyunin_ea17, JimenezSerra_ea21} is designed to simulate chemical evolution under interstellar conditions using a rate equations approach. Similar to MAGICKAL, the MONACO astrochemical code is also a three-phase model that considers the gas phase, the surface of the icy mantle and the bulk of the ice. 
Note that the surface of the icy mantle corresponds to the four uppermost surface layers of the ice.

Chemistry in the gas phase and on grains is coupled via accretion and desorption. Four types of desorption processes are included in the MONACO code: thermal sublimation, photodesorption, cosmic ray-induced desorption and reactive (chemical) desorption. Although several treatments of reactive desorption are included in the code \citep[following][]{Garrod_ea07,VasyuninHerbst13,Minissale2016},  here we employ the approach of \citet[][]{Garrod_ea07}. 

In the code, diffusive and non-diffusive chemical processes on grains are both considered. Diffusive chemistry is implemented following the classical approach by \citet[][]{Hasegawa_ea92} with several modifications. First, the treatment of reaction/diffusion competition is added following \citet[][]{Chang_ea07}. Next, time-dependent adjustment of binding energies of species on the surface due to the presence of H$_2$ molecules is added following \citet[][]{garrodandpauly11}. 
Note, however, that while \citet[][]{garrodandpauly11} adjusted binding energies for all species, we do it only for H and H$_2$. Heavier species are considered to penetrate through the top monolayer of H$_2$ molecules on ice surface during accretion, and bind to the regular ice surface beneath. 
For diffusive chemistry, only thermal hopping is considered as a source of mobility of species. Diffusion-to-binding energy ratios are assumed different for atomic and molecular species. For atoms, $E_{\rm diff}/E_{\rm des}$ = 0.5, while for molecules $E_{\rm diff}/E_{\rm des}$ = 0.3. In the bulk, diffusion energies are twice higher than on the surface for all species except for H and H$_2$. For H and H$_2$, their bulk diffusion energies are 1.5 times higher than values in the ice surface. Tunnelling through reaction activation barriers is allowed. 

Non-diffusive chemistry is considered following the approach proposed by \citet[][]{jinandgarrod20}. Excited three-body reactions are excluded. Ice photochemistry and chemistry induced by cosmic rays is treated as in \citet[][]{Vasyunin_ea17}, although rates of some processes are updated according to recent estimates \citep[see e.g.][]{Jimenezserra2018,JimenezSerra_ea21}.

\subsection{Nautilus code}
\label{sec:nautilus}

Nautilus is an astrochemical gas-grain code based on the rate equation method and available through a git repository\footnote{https://astrochem-tools.org/codes/} \citep[][]{Ruaud2016,wakelam2024}. It is also a three phase model meaning that it takes into account the gas-phase chemistry and it makes a distinction between the few upper layers (four monolayers) of molecules and the rest of the mantles (bulk) on top of interstellar refractory grains. The diffusion of the species is less efficient in the bulk than on the surface. The ratio between the diffusion energies and binding energies is assumed to be 0.8 in the bulk and 0.4 on the surface. Diffusion is assumed to occur via thermal hopping but also through tunnelling depending on the mass of the species. Grains are assumed to be of one size and composition (silicate grains of 0.1 $\upmu$m in radius). Non-diffusive chemistry is not considered in Nautilus. 

In addition to thermal desorption, a number of non thermal desorption processes are included: whole grain heating induced by cosmic-rays \citep{Hasegawa1993}, photodesorption \citep{Ruaud2016}, chemical reactive desorption \citep{Minissale2016}, and sputtering by cosmic-rays \citep[][]{Wakelam2021}. Note that this is the only code in this work that considers the effects of cosmic-rays sputtering as a desorption mechanism following the experimental results of \citet[][]{Dartois2020} and \citet{Dartois2021}. The chemical networks and parameters are the same as in \citet{Clement2023}. 

\subsection{\textsc{Uclchem} code}
\label{sec:uclchem}

\textsc{Uclchem} \citep{Holdship2017} is a time-dependent gas-grain chemical model based on the rate equation approach. \textsc{Uclchem} is a public open source code\footnote{https://uclchem.github.io/.} 
that considers the gas phase as well as the bulk and the surface of dust grains (i.e. it is also a three-phase model). Gas phase reactions are two-body reactions, including interactions with cosmic rays and UV photons. Gas species can accrete onto grain surfaces. Diffusive chemistry is considered as described in \cite{Quenard2018a}, using a constant ratio between diffusion and binding energies of 0.5.  However, \textsc{Uclchem} does not include non-diffusive processes in the ices. Both thermal and non-thermal desorption mechanisms are considered in this code. For thermal desorption, \textsc{Uclchem} follows the formalism of \cite{Viti2004}, while for the non-thermal desorption processes \textsc{Uclchem} includes desorption produced by the energy released in H$_2$ formation, incident cosmic rays, UV photo-desorption, cosmic-ray-induced UV photo-desorption \citep[as described in][]{Holdship2017} and chemical reactive desorption following the formalism of \citet{Minissale2016} \citep[see][]{Quenard2018a}. 
The default grain chemical network is based on the one developed by \cite{Quenard2018a}, and the gas-phase chemical network is taken from the \textsc{Umist} 2013 database \citep{McElroy2013}. The default grain chemical network was updated by adding the new reactions for CH$_3$OH ice formation proposed in Section \ref{sec:parameters} and by changing the activation energy of reaction $\ce{H2CO + O} \,\rightarrow\, \ce{CO2 + H2}$ from 0 to 335 K \citep{Minissale2015}. We also added a set of reactions involving HOCO and \ce{CO2}, which are relevant for the production of \ce{CO2} in the ice (see Appendix \ref{app:uclchem-hoco} and references therein). Note that these reactions were already included in the other codes.

\subsection{Physical parameters and chemical reactions explored by the models}
\label{sec:parameters}

The chemistry of the Chamaeleon I molecular cloud is modelled considering its evolution from its translucent phase to the dense cloud stage. To do this, we first consider a “phase 0” stage in which the chemistry of a translucent cloud is modelled for a duration of $10^{6}$ yr assuming that the cloud is static and adopting a number density of total H nuclei of $n_\mathrm{H}$ = 2$\times10^3$ cm$^{-3}$, a dust and gas temperature $T_\textrm{d}=T_\textrm{gas}=15$ K, visual extinction $A_\textrm{V}=2$\,mag, UV radiation field G$_0$=1 hab, and a cosmic-ray ionisation rate $\zeta$=1.3$\times$10$^{-17}$ s$^{-1}$. 

The chemical evolution of the cloud collapse is modelled in a subsequent "phase 1" stage, in which the output gas and ice (both surface and bulk) abundances from ``phase 0'' are used as input abundances for the ``phase 1'' simulations. For the cloud collapse, we assume a free-fall collapse treatment with the gas density ($n_\mathrm{H}$) evolving with time ($t$) as \citep{brown88,rawlings1992}:

\begin{equation}
    \frac{\mathrm{d}\,n_\mathrm{H}}{\mathrm{d}\,t} \;\;=\;\; b\; \left ( \frac{n_\mathrm{H}^4}{n_\mathrm{H,0}} \right )^\frac{1}{3} \left [ 24\uppi \, G \, m_\mathrm{H} \, n_\mathrm{H,0}  \left ( \left ( \frac{n_\mathrm{H}}{n_\mathrm{H,0}}  \right ) ^\frac{1}{3} - 1 \right ) \right ] ^\frac{1}{2}  \; .
\label{free-fall}
\end{equation}

$n_\mathrm{H,0}$ is the initial hydrogen nuclei central density, $m_\mathrm{H}$ is the mass of a hydrogen nucleus, and $G$ is the gravitational constant. We also include a factor $b$ in the equation so that we can test the effects of an accelerated collapse in the chemistry of the ices  \citep{Holdship2017}. The collapse finishes when $n_\mathrm{H}$ reaches the final density values at the final collapse times shown in Table$\,$\ref{tab:gridofmodels}. As the density ${n_\textrm{H}}$ evolves during the collapse, the visual extinction $A_\textrm{V}$ increases as a function of density (and, therefore, time), following Equation 1 of \citet{Jimenezserra2018}. 

Table \ref{tab:gridofmodels} reports the grid of physical parameters considered for the models of the two field stars NIR38 and J110621. The assumed visual extinctions are $A_\mathrm{V}$ = 60 mag and $A_\mathrm{V}$ = 95 mag, respectively; however, these values may be lower \citep[see][and Section \ref{sec:Av}]{dartois2024}. The considered gas/dust temperatures are 14 K for NIR38 and 12 K for J110621, as inferred from the Herschel-based dust temperatures maps of \citet{Vaisala2014}\footnote{The uncertainties associated with these T$_{\rm dust}$ values are as high as $\pm$0.6 K, after considering a $\beta$ range between 1.8 and 2.2 and the uncertainties due to photometric noise \citep{Vaisala2014}.}. For the MAGICKAL and MONACO codes, the dust temperature is assumed to decrease from 15 K during the core collapse with increasing visual extinction as described in \citet{garrodandpauly11}. If the temperature reaches the minimum value of the line of sight (12~K or 14~K depending on the field star; Table \ref{tab:gridofmodels}), the temperature is then fixed. For the rest of the codes, $T_\textrm{dust}$ is kept constant at either 12 K or 14 K (see Table \ref{tab:gridofmodels}). The gas temperature is set to be the same as the dust temperature in all runs and for all codes.

The final $n_{\rm H}$ densities at the end of the "phase 1" collapse stage are 2$\times$$10^4$, 2$\times$$10^5$ and 2$\times$$10^6$ cm$^{-3}$. These densities have been selected from the \ch{H2} volume gas densities estimated by \citet{Belloche11} toward the Cha1-C1 and Cha1-C2 cores (see their Table 6), which are consistent with the results of \citet{Vaisala2014}. These densities lie between some $10^5$ and 10$^6$ cm$^{-3}$. For completeness, we also explore the lower-density case with $n_{\rm H}$ = 2$\times$$10^4$ cm$^{-3}$.

For the final collapse times, we assume time-scales of $10^4$, $10^5$ and $10^6$ yr. Free-fall collapse times\footnote{Free-fall collapse times are calculated as $t_{\rm free-fall}\simeq 4\times 10^7/\sqrt{n_{\rm H,0}}$\,yr, with n$_{\rm H,0}$ the initial gas cloud density.} correspond to $\sim$10$^6$ yr for an initial gas cloud density of n$_{H,0}$=2$\times$10$^3$$\,$cm$^{-3}$ (i.e. the one used in the "phase 0" stage of the simulations). The collapse times of $10^4$ and $10^5$ yr are tested by modifying the $b$ parameter in Equation \ref{free-fall}, which adopts values from 1 (free-fall, $\sim10^6$ yr) to 120 (accelerated collapse, $\sim10^4$ yr). The three different collapse times assumed in our grid of models are included to test the effects of a fast/slow collapse on the chemical composition of the ices. Note that short time-scales of the order of 10$^4$ yr are typically not observed for prestellar cores \citep{keto2010}. However, accelerated collapse is considered in this study to investigate the possible effects of gravitational collapse driven by stellar feedback from nearby star formation\footnote{A 10 km s$^{-1}$ shock propagating through a preshock gas with density of $\sim$10$^{4}$ cm$^{-3}$ would induce a density compression in the postshock gas by only a factor $\sim$3 for time scales $\sim$10$^4$ yrs \citep[see][]{jimenez2008,gusdorf2008}.}. We note that we deliberately do not explore a larger grid of models because the KMC simulations are computationally very expensive. The cosmic-ray ionisation rate and UV radiation field assumed for the "phase 1" stage in all models are, respectively, $\zeta$=1.3$\times$10$^{-17}$ s$^{-1}$ and G$_0$=1 hab.

As initial elemental abundances, we consider the low-metal abundance case (the $"$EA1$"$ case) from \citet[][see Table$\,$\ref{tab:initial}]{Wakelam08}. 
All models consider a constant grain size of 0.1 $\upmu$m. The output fractional abundances are given with respect to the number density of total H nuclei. The conversion from ice molecular abundances to ice molecular column densities is performed using the column densities of total H nuclei toward NIR38 and J110621 calculated\footnote{This expression differs only by a factor of $\sim$1.1 with respect to the one reported by \citet{lacy2017} of $N_{\rm H} \,=\, 1.8 \times 10^{21} \, A_{\rm V} \, {\rm cm}^{-2}$.} as $N_{\rm H} \,=\, 1.6 \times 10^{21} \, \times \, A_{\rm V} \, {\rm cm}^{-2}$.

For the formation of methanol, besides the classical route of multiple CO hydrogenation reactions in the ice, we also include the following chemical processes: 1) the chain of neutral-neutral grain surface reactions CH$_4$ + OH $\rightarrow$ CH$_3$ + H$_2$O and CH$_3$ + OH $\rightarrow$ CH$_3$OH proposed by \citet{Qasim18}; and 2) the reaction C + H$_2$O $\rightarrow$ H$_2$CO and subsequent hydrogenation reactions proposed by \citet{molpeceres2021carbon} and \citet{Potapov21}. 

\begin{table*}
\setlength{\tabcolsep}{10pt} 
\caption{Physical parameters assumed for the grid of models.}\label{tab:gridofmodels} 
\centering           
\begin{tabular}{lcccc|cccc}  
\hline   
& \multicolumn{4}{c}{NIR38} & \multicolumn{4}{c}{J110621} \\ \cline{2-9}
Model & $n_{\rm H}$ & $T_{\rm dust}$ & $A_{\rm V}$ & Collapse time & $n_{\rm H}$ & $T_{\rm dust}$ & $A_{\rm V}$ & Collapse time \\
& (\rm cm$^3$) & (\rm K) & (\rm mag) & (\rm yr) & (\rm cm$^3$) & (\rm K) & (\rm mag) & (\rm yr) \\
\hline         
1 & 2$\times$10$^4$ & 14 & 60 & 10$^4$ & 2$\times$10$^4$ & 12 & 95 & 10$^4$ \\
2 & 2$\times$10$^4$ & 14 & 60 & 10$^5$ & 2$\times$10$^4$ & 12 & 95 & 10$^5$ \\
3 & 2$\times$10$^4$ & 14 & 60 & 10$^6$ & 2$\times$10$^4$ & 12 & 95 & 10$^6$ \\
4 & 2$\times$10$^5$ & 14 & 60 & 10$^4$ & 2$\times$10$^5$ & 12 & 95 & 10$^4$ \\
5 & 2$\times$10$^5$ & 14 & 60 & 10$^5$ & 2$\times$10$^5$ & 12 & 95 & 10$^5$ \\
6 & 2$\times$10$^5$ & 14 & 60 & 10$^6$ & 2$\times$10$^5$ & 12 & 95 & 10$^6$ \\
7 & 2$\times$10$^6$ & 14 & 60 & 10$^4$ & 2$\times$10$^6$ & 12 & 95 & 10$^4$ \\
8 & 2$\times$10$^6$ & 14 & 60 & 10$^5$ & 2$\times$10$^6$ & 12 & 95 & 10$^5$ \\
9 & 2$\times$10$^6$ & 14 & 60 & 10$^6$ & 2$\times$10$^6$ & 12 & 95 & 10$^6$ \\ \hline
\end{tabular}
\tablefoot{$T_{\rm gas}$ is assumed to be the same as $T_{\rm dust}$ in all models.}
\end{table*}

\begin{table}
\setlength{\tabcolsep}{10pt} 
\caption{Initial elemental abundances with respect to n$_{\rm H}$}\label{tab:initial} 
\centering           
\begin{tabular}{lc}  
\hline   
Species & Abundance \\ \hline
He & 1.40$\times$10$^{-1}$ \\
N & 2.14$\times$10$^{-5}$ \\
O & 1.76$\times$10$^{-4}$ \\
C & 7.30$\times$10$^{-5}$ \\ \hline
\end{tabular}
\tablefoot{Abundances correspond to the $"$EA1$"$ low-metal abundance case of \citet{Wakelam08}. These values consider the partial depletion of these elements into dust grains.}
\end{table}

\section{Results} 
\label{sec:results}

In this Section, we analyse the results obtained with each code, presenting the evolution of the chemical composition of the ices and the main processes that lead to their formation for the grid of models described in Section \ref{sec:parameters}. We focus on some of the molecular species observed with JWST toward NIR38 and J110621, namely H$_2$O, CO, CO$_2$, CH$_3$OH, NH$_3$ and CH$_4$. The model results are thus compared to the observational data reported in \citet{McClure2023} and presented in Table \ref{tab:observedices}. In Section \ref{sec:phase0}, we compare the output from all models for their phase 0 runs to identify any possible difference arising from this stage of the simulations. In Sections \ref{sec:results-magickal}, \ref{sec:results-monaco}, \ref{sec:results-nautilus}, \ref{sec:results-uclchem} we present the results for the phase 1 runs from the rate equation codes MAGICKAL, MONACO, Nautilus and \textsc{Uclchem}, respectively, while in Section \ref{sec:results-kmc}, we report the results from the stochastic KMC model. 

\begin{table*}
\setlength{\tabcolsep}{10pt} 
\caption{Observed column densities and column density ratios with respect to H$_2$O toward NIR38 and J110621 \citep{McClure2023}}\label{tab:observedices} 
\centering
\renewcommand*{\arraystretch}{1.2}
\begin{tabular}{lcccc}  
\hline    
& \multicolumn{2}{c}{NIR38} & \multicolumn{2}{c}{J110621} \\
& \multicolumn{2}{c}{($A_{\rm V}$ = 60 mag)} & \multicolumn{2}{c}{($A_{\rm V}$ = 95 mag)} \\ \cline{2-3} \cline {4-5}
Molecule & $N$(molecule) &  $N$(molecule)$/$$N$(H$_2$O) & $N$(molecule) & $N$(molecule)$ / $$N$(H$_2$O) \\
& (\rm $10^{18}$ cm$^{-2}$) & (in \%) & (\rm $10^{18}$ cm$^{-2}$) & (in \%) \\
\hline         
H$_2$O &	$6.9_{-1.1}^{+1.9}$ &	100 &	$13.4_{-1.8}^{+1.3}$ &	100 \\
CO	& $3.0_{-0.4}^{+0.6}$ & $43_{-10}^{+12}$ &	$3.7_{-0.4}^{+0.6}$ & $28_{-4}^{+6}$ \\
CO$_2$ &	$1.38_{-0.20}^{+0.20}$ & $20_{-5}^{+5}$ &	$1.74_{-0.22}^{+0.21}$ & $13.1_{-2.0}^{+2.5}$ \\
CH$_3$OH &	$0.61_{-0.11}^{+0.11}$ & $8.7_{-2.1}^{+2.5}$ &	$0.51_{-0.09}^{+0.19}$ & $3.9_{-0.8}^{+1.5}$ \\
NH$_3$ &	$0.30_{-0.03}^{+0.22}$ & $4.6_{-1.1}^{+3.0}$ & $0.66_{-0.06}^{+0.15}$ & $5.0_{-0.7}^{+1.3}$ \\
CH$_4$ &	$0.180_{-0.013}^{+0.017}$ &	$2.6_{-0.6}^{+0.5}$ &	$0.250_{-0.030}^{+0.010}$ &	$1.84_{-0.24}^{+0.30}$ \\ \hline
\end{tabular}
\renewcommand*{\arraystretch}{1.0}
\end{table*}

\subsection{Comparison of the Phase 0 results from all models}
\label{sec:phase0}

\begin{figure}
\centering
\includegraphics[width=1.0\linewidth]{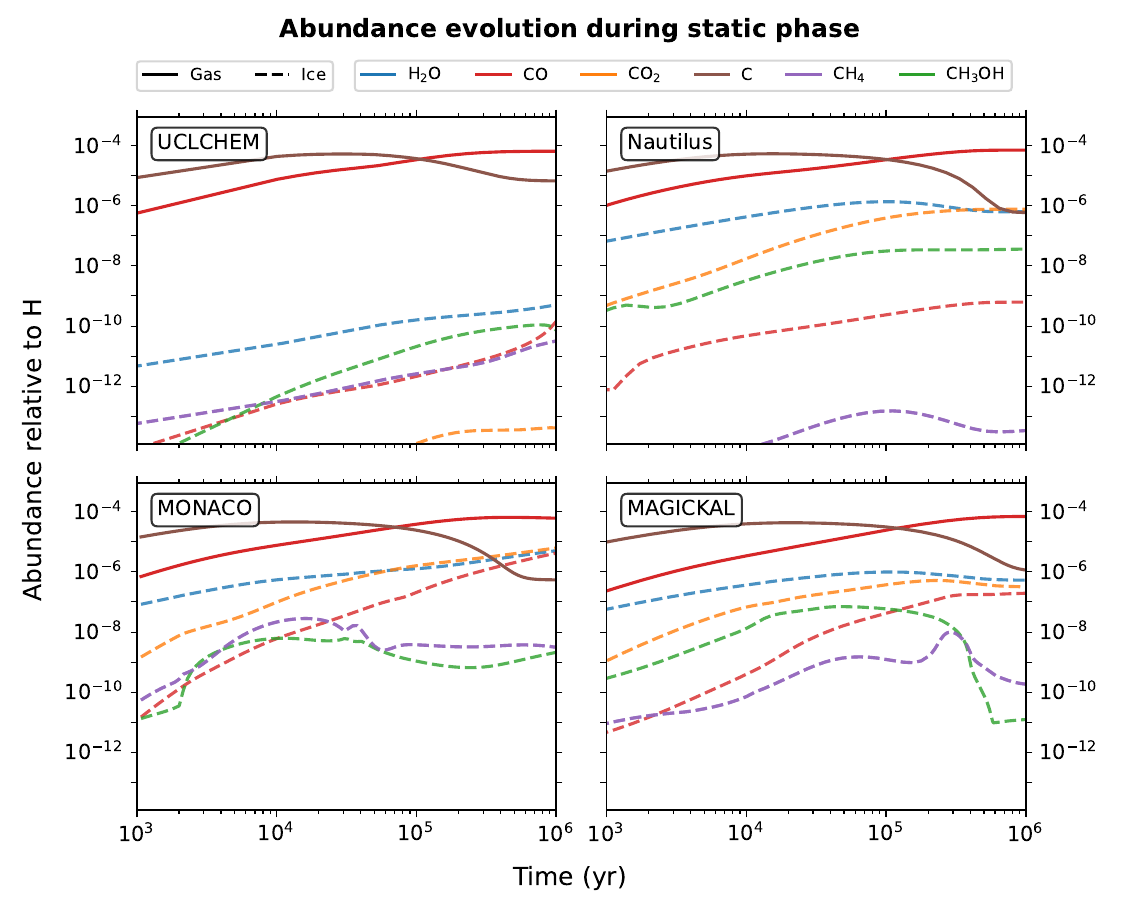}
\caption{Time evolution of the gas and ice molecular abundances for most relevant species predicted by the different astrochemical codes for the phase 0 runs. Solid lines indicate gas-phase abundances, while dashed lines show ice abundances. Colours indicate the different molecular species as shown in the upper part of the figure.}
\label{figure-staticphase}
\end{figure}

Figure \ref{figure-staticphase} shows the abundances with respect to \ce{H2} resulting from the phase 0 stage of the simulations, where we report the time evolution of the relevant species modelled by the different astrochemical codes. For the gas-phase, we only plot the relative abundances of atomic C and CO since these two species are essential to understanding the amount of C that will be locked into CO, CO$_2$, and CH$_4$ ices during the phase 1 of the simulations. Note that we do not show the phase 0 results for the KMC model, because they are similar to the ones from the \textsc{Uclchem} code. The largest differences found between the \textsc{Uclchem} and KMC phase 0 results are the abundances of \ce{CO2} for which several reactions have been added within \textsc{Uclchem} (see Appendix \ref{app:uclchem-hoco}).  

From Figure \ref{figure-staticphase}, it is clear that for all models atomic C has been efficiently converted into CO by the end of the phase 0 stage with a CO gas-phase abundance that goes up to $\sim$$7\times 10^{-5}$ with respect to the total H nuclei at the end of the simulations. As shown in Sections \ref{sec:results-magickal}, \ref{sec:results-monaco}, \ref{sec:results-nautilus}, \ref{sec:results-uclchem}, this will largely affect the subsequent production of CO$_2$ and CH$_4$ during the phase 1 stage of the models. For the Nautilus and MAGICKAL codes, the temporal evolution of the gas and ice molecular abundances are similar, although some differences can be seen for CH$_3$OH and CH$_4$. In these models, H$_2$O and CO$_2$ are the most abundant ice species with relative abundances lying between $10^{-7}$ and $10^{-6}$. Due to the dust temperature of $T_{\rm dust} $= 15~K considered for the phase 0 runs, any CO molecule sticking on the grains is converted into CO$_2$, 
which explains the high abundance of CO$_2$ ice (even higher than that of CO) for these models. For the MONACO code, the final abundances of the phase 0 runs are somewhat higher than for the other codes (see Figure~\ref{figure-staticphase}), with an ice thickness of $\sim$10 monolayers, versus $\sim$1 monolayer or less.

For the \textsc{Uclchem} code, Figure \ref{figure-staticphase} shows that the ices grow at a slower effective rate compared to the other codes. This is produced by the higher efficiency of non-thermal desorption in \textsc{Uclchem} as compared to other codes. In particular, UV-photodesorption is the mechanism responsible for the slow net ice growth at these low visual extinctions ($A_\text{V}$=2 mag). We note, however, that if the assumed UV yield in \textsc{Uclchem} (i.e. the number of molecules desorbed per UV photon) is reduced from 0.1 molecules photon$^{-1}$ to 0.03 molecules photon$^{-1}$, the predicted ice abundances become similar to those given by the other codes. For the remaining of this work, we use the default value of 0.1 molecules photon$^{-1}$ given in \textsc{Uclchem}, to evaluate the effects of a less efficient ice growth during the static phase 0 stage in a translucent cloud (see Section \ref{sec:results-uclchem}). 

In the following sections, we present the results from the phase 1 calculations: first, for all rate equations codes (presented in alphabetical order); and second, for the KMC code.  

\subsection{Collapse Phase 1 results}
\label{sec:results-phase1}

\subsubsection{Results from the MAGICKAL code}
\label{sec:results-magickal}

\begin{figure*}
\centering
\includegraphics[width=0.85\linewidth]{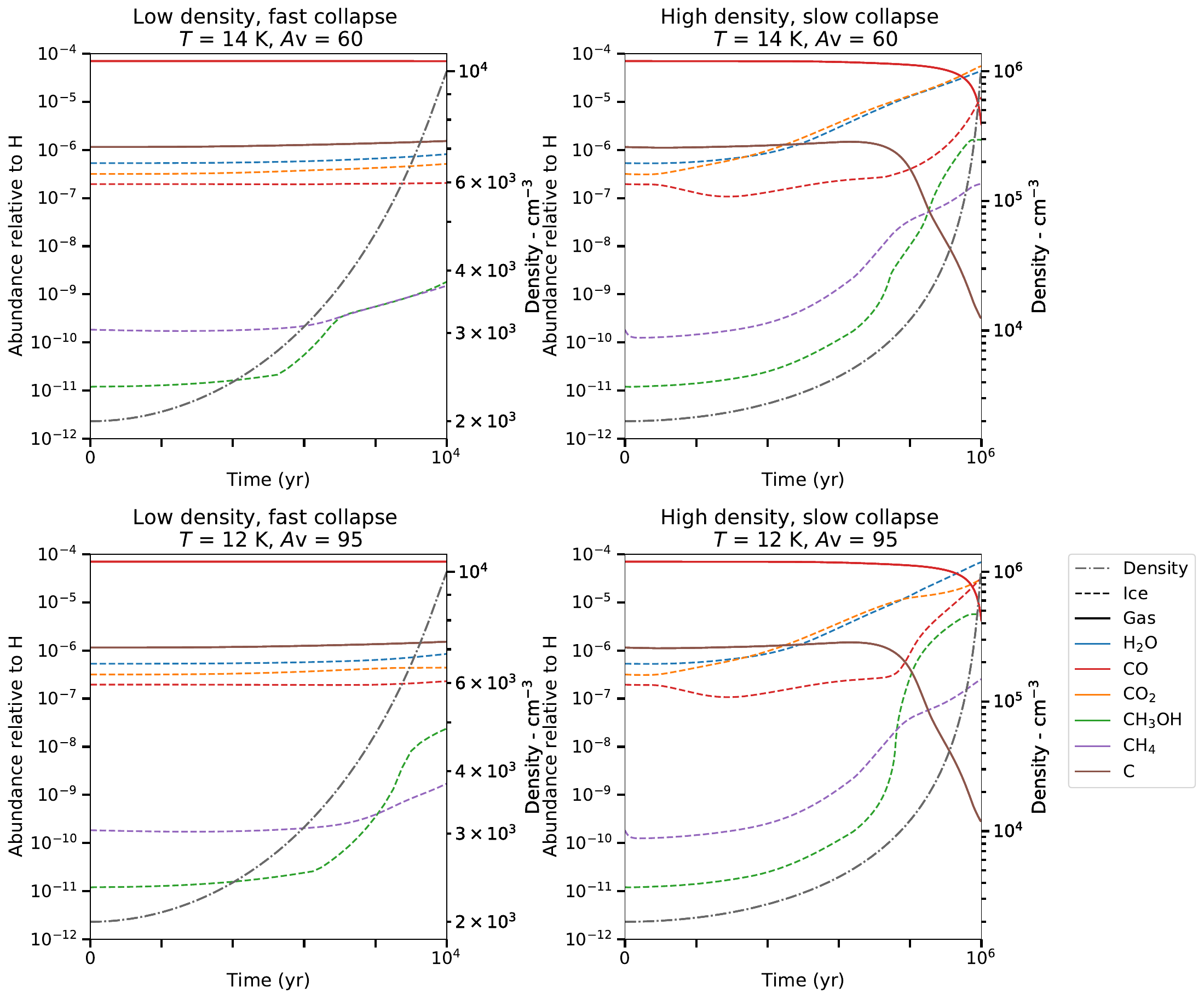}
   \caption{The time evolution of gas (solid) and ice (dashed) abundances in \textit{MAGICKAL} during the collapse $"$phase 1$"$ stage. Models with T$_{\rm dust}$=14 K and $A_{\rm v}$=60 mag correspond to NIR38, while models with T$_{\rm dust}$=12 K and $A_{\rm V}$=95 mag  correspond to J110621.} \label{magickal-collapse}
\end{figure*}

In Figure \ref{magickal-collapse}, we plot the time evolution of the gas and ice relative abundances of H$_2$O, CO, CO$_2$, CH$_3$OH, CH$_4$ and atomic C during the phase 1 stage of the collapse, for four representative models obtained with the MAGICKAL code. As examples, we report  the most extreme cases with the lowest-density and fastest-collapse models (models \#1; left panels) and the highest-density and slowest-collapse models (models \#9; right panels) for NIR38 and J110621, to provide information about the maximum changes expected within the grid of models.
In the low-density and fast-collapse models (left panels, Figure \ref{magickal-collapse}), the chemical evolution is nearly identical for NIR38 and J110621 regardless of temperature, except for \ce{CH3OH} ice. The gas phase abundances of CO remain largely unchanged over time due to the collapse timescale being much shorter than the accretion timescale of CO. Consequently, this leads to almost constant abundances of CO ice and thus \ce{CO2}, as \ce{CO2} primarily forms from the association of CO and OH on the surface. However, in the model of J110621 with T$_{\rm dust}$=12 K, hydrogenation becomes more efficient, resulting in the formation of more \ce{CH3OH} via the hydrogenation of CO. 

In the models with high density and slow collapse (right panels; Figure \ref{magickal-collapse}), gas-phase CO and C efficiently deplete on grains as density evolves, significantly impacting surface chemistry. As can be seen from this figure, the abundance of CO ice remains relatively constant (or even decreases) until the density of the gas becomes high enough for the catastrophic depletion of CO to take place. Surface CO then rapidly reacts forming primarily \ce{CO2} through the grain surface reaction CO + OH $\rightarrow$ CO$_2$ + H \citep[see][]{garrodandpauly11,Clement2023}, where OH forms via the hydrogenation of adsorbed atomic oxygen. Once \ce{CO2} is formed, it accumulates in the ice due to its chemically inert characteristics. Consequently, the ice abundance of \ce{CO2} in these models gradually increases over time, while CO does not, resulting in a \ce{CO2} / CO ratio $>$ 1. The abundance of \ce{CH3OH} ice exhibits an evolutionary trend similar to that of CO since it primarily forms from the hydrogenation of CO on the surface. In the model with T$_{\rm dust}$=12 K, methanol is more abundant due to the higher abundance of CO on the grain surface and the higher efficiency of CO hydrogenation. The contribution of the reactions CH$_3$ + OH and C + H$_2$O to the production of methanol in these models is minor compared to grain-surface CO hydrogenation. This is due to the fact that atomic C is mainly converted into gas-phase CO, which quickly freezes out onto dust grains to form either CH$_3$OH or CO$_2$. This leaves little atomic C in the gas phase to adsorb on grains and subsequently hydrogenate forming CH$_4$, or to react with H$_2$O in the ice forming H$_2$CO (see also Sections$\,$\ref{sec:ch3oh} and \ref{sec:hydrides}).

\begin{figure*}
\centering
\includegraphics[width=1\linewidth]{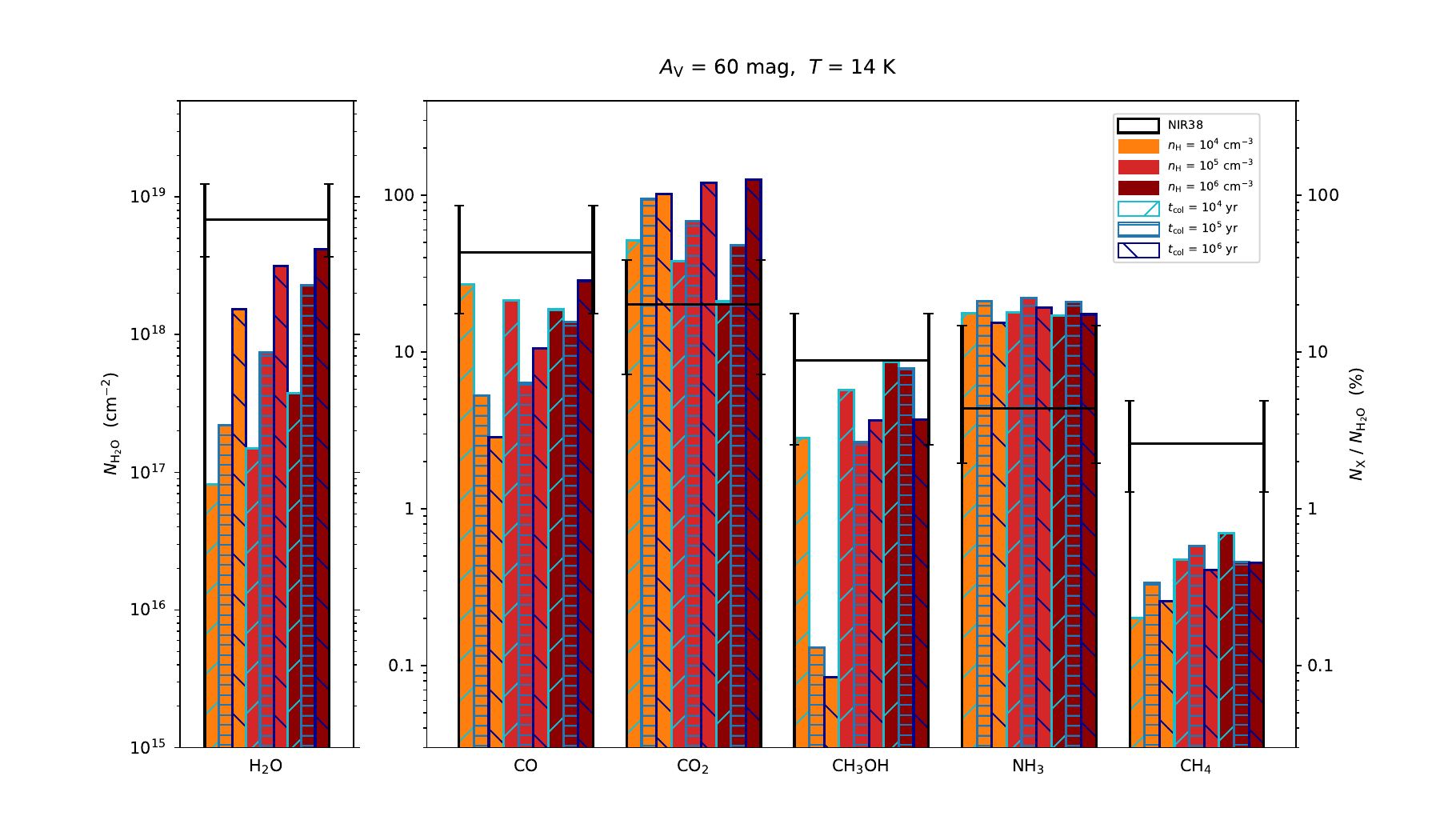}
\caption{Ice compositions predicted by {\it MAGICKAL} for NIR38 for the models listed in Table \ref{tab:gridofmodels}. Left panel: Predicted water ice column densities. Wide right panel: Ice column density ratios with respect to water for the remaining species. Histograms are coloured and drawn as indicated in the upper right part of the panel. The visual extinction and dust temperatures assumed in the models are reported in the upper part of the panel. Black wide histograms report the observed Ice Age column density of water and ice column density ratios obtained by \citet[][]{McClure2023}. The 3$\sigma$ error bars are also shown. \label{magickal-results-Av60}}
\end{figure*}

\begin{figure*}
\centering
\includegraphics[width=1\linewidth]{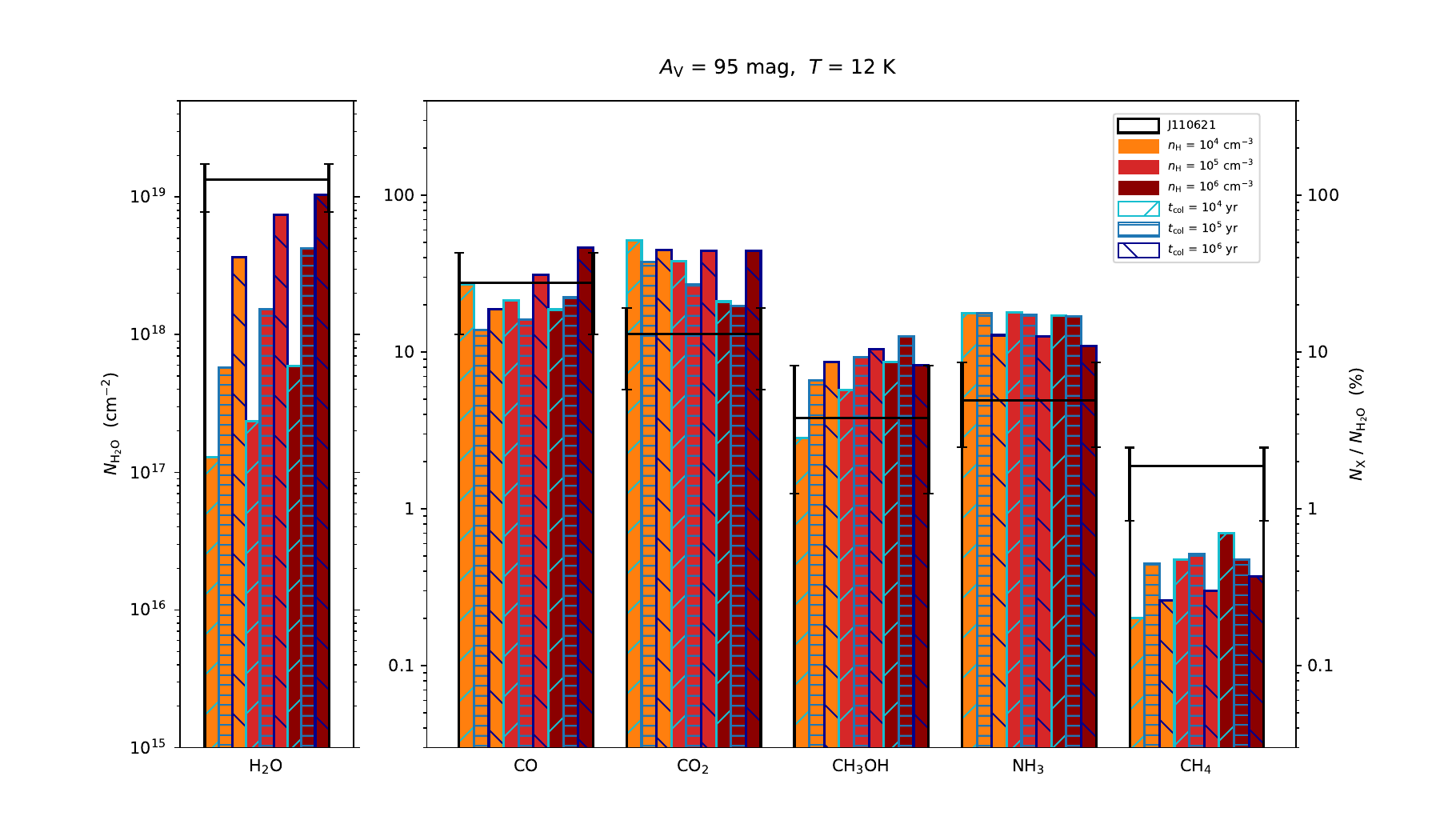}
\caption{Predicted ice compositions by {\it MAGICKAL} for J110621. Details are as for Figure \ref{magickal-results-Av60}.    \label{magickal-results-Av95}}
\end{figure*}

The final ice chemical compositions toward NIR38 and J110621 for the whole grid of models can be found in 
Figures~\ref{magickal-results-Av60} and \ref{magickal-results-Av95}. In these figures, we provide the absolute column densities of H$_2$O predicted in the ice at the end of the phase 1 simulations, together with the ice abundance ratios of the rest of species analysed in this study (CO, CO$_2$, CH$_3$OH, NH$_3$ and CH$_4$) with respect to H$_2$O in \%. From Figures~\ref{magickal-results-Av60} and \ref{magickal-results-Av95}, we find that, regardless of the model used, both lines of sight consistently underpredict water ice column densities and the ice abundances of \ce{CH4}. However, the water ice column density of the model with the longest collapse timescale ($t_\textrm{col}=10^6$ yr) and the highest density ($n_\textrm{H}=10^6~\textrm{cm}^{-3}$) aligns with observations within the margin of error. This is due to the fact that models with longer collapse timescales enable the accumulation of larger amounts of H$_2$O in the ice mantle. CO, \ce{CO2} and \ce{CH3OH} are the most sensitive species to variations depending on the model parameters. Ice abundances of \ce{CO2} surpass observational values in all models for both lines of sight. CO$_2$ becomes more abundant in the ice than even CO yielding \ce{CO2} / CO ratios $\geq$1, which is not observed toward either the field star NIR38 or J110621. For \ce{NH3} and \ce{CH4}, we find similar ice abundances across all models given that both species are formed through the consecutive hydrogenation of atomic N and C. 

To summarise, despite the observed differences among models, the ones that  better match the observations are those with higher densities and longer collapse timescales. In Section \ref{sec:best}, we will evaluate the model that best matches the ice chemical compositions observed within Ice Age.

\subsubsection{Results from the MONACO code}
\label{sec:results-monaco}

The evolution of the ice chemical composition during the collapse (phase 1) stage obtained with the MONACO code is presented in Figure~\ref{figure-MONACO-collapse} in Appendix \ref{sec:evolution}. 
The evolution of the ice molecular abundances along the collapse is similar to that predicted by the MAGICKAL code.
The final ice compositions for all considered models with the MONACO code are presented in Figures~\ref{monaco-results-Av60} and \ref{monaco-results-Av95}. 

As can be seen from these Figures, the abundances of major ice species (H$_2$O, CO, CO$_2$) behave differently in the models with T$_{\rm dust}$~=~14~K from those with T$_{\rm dust}$~=~12~K. At T$_{\rm dust}$~=~14~K, solid CO is converted into CO$_2$ ice more efficiently than at T$_{\rm dust}$~=~12~K. The major route for CO$_2$ ice formation is the reaction CO~+~OH~$\rightarrow$~CO$_2$~+~H. At  T$_{\rm dust}$~=~14~K, this reaction proceeds more efficiently both due to the higher probability to overcome the reaction activation barrier and also due to the non-zero mobility of CO molecules at this dust temperature. Interestingly, the contribution of the surface reaction O~+~CO~$\rightarrow$~CO$_2$ to the formation of CO$_2$ is minor in the MONACO model both at T$_{\rm dust}$~=~14~K and T$_{\rm dust}$~=~12~K. Solid H$_2$O, on the contrary, forms more efficiently on colder grains, which results in twice higher final abundance of H$_2$O ice at 12~K. The abundance of solid CO$_2$ surpasses that of solid H$_2$O at 14 K, despite the continuous H$_2$O ice accumulation. At 12~K, however, the result is the opposite: H$_2$O is the most abundant ice species, with a predicted solid abundance that is a factor of $\sim$3 higher than those of CO and CO$_2$ ices.

From Figures \ref{monaco-results-Av60} and \ref{monaco-results-Av95}, we find that the fast-collapse models ($t_{\rm col}$=10$^4$ years) exhibit the worst agreement with the Ice Age observations both in terms of absolute H$_2$O ice column densities and ice chemical composition. The models of NIR38 significantly overproduce solid CO$_2$, yielding abundances that clearly exceed those of H$_2$O ice, which is not observed \citep{McClure2023}. The models of J110621, with lower dust temperatures (T$_{\rm dust}$~=~12~K), produce in general more reasonable results. For the J110621 runs, CO and CO$_2$ are produced in similar amounts (i.e. 20--40\% the abundance of H$_2$O ice) in models with intermediate/slow collapse times ($t_{\rm col}$=10$^5$ and 10$^6$ yrs) and intermediate/high final densities (2$\times$10$^5$ and 2$\times$10$^6$ cm$^{-3}$). However, for the slow-collapse models ($t_{\rm col}$=10$^6$ yrs), the CO/CO$_2$ ice abundance ratio reaches values $\sim$2, which is similar to the one measured with the Ice Age data. We note that the rest of models results in CO/CO$_2$ abundance ratios $<$1. 

Methanol ice is severely underproduced in the models with T$_{\rm dust}$~=~14~K, which can be related to the overproduction of CO$_2$ ice (Figures \ref{monaco-results-Av60} and \ref{monaco-results-Av95}). The formation of CH$_3$OH in the MONACO model runs occurs via successive hydrogenation of CO, which is driven by diffusive H-addition reactions. No significant impact on the final CH$_3$OH ice abundance is observed when including the chain of grain-surface reactions CH$_4$~+~OH and C~+~H$_2$O proposed by \citet{Qasim18}, \citet{molpeceres2021carbon} and \citet{Potapov21}. The final abundance of CH$_3$OH ice is higher in the model with T$_{\rm dust}$~=~12~K as expected from the low binding energy of atomic hydrogen. 
As for CH$_3$OH, CH$_4$ ice is also underproduced in all models, although the slow-collapse models tend to provide better agreement with the Ice Age observations.

In summary, models with higher final densities and collapse timescales of 10$^5$--10$^6$ yrs provide a closer match to the ice chemical composition obtained by the Ice Age observations. This conclusion is similar to those obtained for the MAGICKAL chemical code (see Sections \ref{sec:results-magickal} and \ref{sec:results-nautilus}).

\subsubsection{Results from the Nautilus code}
\label{sec:results-nautilus}

The evolution of the ice and gas phase abundances obtained by the Nautilus code for the phase 1 stage is reported in the Appendix (Figure \ref{figure-nautilus-collapse}). The evolution of the ice molecular abundances is similar to that predicted by the MAGICKAL and MONACO codes as described in Appendix \ref{sec:evolution}.

The predicted column densities of water ice and the ice abundance ratios obtained by the Nautilus code for phase 1 for the different collapse models are shown in Figures~\ref{nautilus-results-Av60} and \ref{nautilus-results-Av95}. As for the MAGICKAL and MONACO codes, CO is formed in the gas-phase and subsequently depletes onto the surface of dust grains. The CO ice molecular abundances increase with density and time, although not linearly. Because of the dust temperatures $T_{\rm dust}$$\geq$12~K, a large fraction of the CO that sticks on the grains is converted into either CH$_3$OH or CO$_2$. For CH$_3$OH, the main formation process is hydrogenation of CO. However, for CO$_2$, its conversion from CO does not occur via the reaction CO + OH $\rightarrow$ CO$_2$ + H as found for the MAGICKAL and MONACO codes, but through the grain surface reaction CO + O $\rightarrow$ CO$_2$. This conversion is very efficient, and the fraction of CO ice is clearly underestimated while the CO$_2$ and CH$_3$OH fractions tend to be overestimated with respect to the Ice Age observations, independently of the n$_{\rm H}$ gas density and the collapse timescales considered in the models. However, note that the ice abundances of CO increase for the highest-density, slowest collapse models, which induce a decrease in the predicted CO$_2$ ice abundances. As found for the MAGICKAL and MONACO codes, the predicted ice abundances get closer to the observed values for the latter models, and hence they are the ones that best reproduce the Ice Age observations. 

For NH$_3$ and CH$_4$, these species are mostly formed by the successive hydrogenation of N and C on the grain surfaces. While our models do a good job at reproducing the measured ice abundances of NH$_3$, CH$_4$ is systematically underproduced in all cases except for the slow-collapse models with $t_\textrm{col}$=10$^6$ yrs (see Figures ~\ref{nautilus-results-Av60} and \ref{nautilus-results-Av95}).

\subsubsection{Results from the UCLCHEM code}
\label{sec:results-uclchem}

In Figure \ref{figure-uclchem-collapse}, we present the evolution of the relevant species both in the ice and in the gas phase for the phase 1 runs (collapse stage). The growth of the ices mainly occurs at the end of the collapse phase with the rapid increase of the density. The rate of ice growth in the \textsc{Uclchem} models is, however, higher than for the previous codes (MAGICKAL, MONACO and Nautilus) because the initial abundances for the phase 1 runs are lower as a result of the efficient non-thermal desorption during the static (phase 0) translucent cloud stage (see Section \ref{sec:collapse}). This is likely the reason for the underproduction of water ice in the \textsc{Uclchem} runs, although the predicted values for some of the models still lie within a factor of 10 with respect to the ice abundances observed with the JWST (see Figures \ref{figure-uclchem-results-T14} and \ref{figure-uclchem-results-T12}). 

For CO, $\ch{CO2}$ and $\ch{CH3OH}$, the predicted ice abundances are either underestimated or overestimated depending on the model. Like for the other codes, the overproduction of CO$_2$ is detrimental for CH$_3$OH formation as the majority of atomic C is locked into the inert CO$_2$ species. According to our models, the main formation route of \ce{CH3OH} on grains at the high visual extinctions of the field stars NIR38 and J110621 is the hydrogenation of CO on dust grain surfaces. Regarding \ce{CO2}, its formation mainly occurs on the surface of dust grains via the reaction CO + OH $\rightarrow$ H + \ce{CO2}. 
We find no difference in the final ice abundance of CO, \ce{CO2} and \ce{CH3OH} between the case of adding the reactions C + \ce{H2O} $\rightarrow$ \ce{H2CO}, and CH$_4$ + OH $\rightarrow$ CH$_3$ + H$_2$O and CH$_3$ + OH $\rightarrow$ CH$_3$OH to the chemical network and the case of not adding them.

For $\ch{NH3}$ and $\ch{CH4}$, \textsc{Uclchem} tends to overpredict the ice abundance of these molecules by factors $\sim\,$2--8. It is interesting to note that \textsc{Uclchem} provides the best match to the observed ice $\ch{CH4}$ abundance. This is due to the fact that gas-phase atomic C is roughly one order of magnitude more abundant in the \textsc{Uclchem} runs than in the runs for the other codes (see Figure \ref{figure-uclchem-collapse}), which yields the depletion of larger amounts of atomic C onto dust grains, enhancing the production of $\ch{CH4}$ ice. 

Overall, the models that best reproduce the observed ice compositions are those with intermediate collapse times ($t_{\rm col} = 10^5\, {\rm yr}$) and intermediate or high final densities ($n_{\rm H} = 10^5, 10^6\, {\rm cm}^{-3}$) (see Section \ref{sec:best}), although there are larger mismatches for the field star J110621 (see Figures \ref{figure-uclchem-results-T14} and \ref{figure-uclchem-results-T12}).   

\subsubsection{Results from the KMC model}
\label{sec:results-kmc}

Figure \ref{figure-KMC-collapse} shows the evolution of the ice chemical composition during the phase 1 runs for the low-density, fast-collapse models and for the high-density, slow-collapse models. The time-dependent gas abundances of CO and C -- modelled by UCLCHEM gas-phase simulations -- are also included in Fig.~\ref{figure-KMC-collapse}. The high initial CO abundance compared to C ($10^{-4}$ versus $10^{-6}$ with respect to total H nuclei) drives the CO-based chemistry on the grain. With fast collapse and low density, no significant freeze-out of CO occurs, while with the slow-collapse high-density models, depletion of CO starts at later times ($>$10$^{5}$ yrs) induced by the rapid increase of density. A steady increase of ice (Bulk) abundances is generally observed throughout the collapse. We note that the increase rate seems unaffected by the accelerated increase of density, except for CO in the slow collapse models, when the above-mentioned freeze-out occurs. 

The heavy fluctuations observed for \ce{CH4} is a finite-size effect, due to the very low concentration numbers generated by the KMC on the grain. That is, the \ce{CH4} abundance is found to fluctuate between 1 and 0 on our $50\times 50$ lattice. Hence, not much value should be given to the variation of \ce{CH4} abundance shown in this Figure. Note that the $50\times50$ size surface model is chosen as a compromise between simulation time and size effect. The algorithm scales at least with $N^2 \log N$ and hence much larger sizes become prohibitively expensive. Fortunately, because of the large number of species on the grain the finite size effect is rather limited and indeed a few short test simulations show that the  $100\times100$ size grains lead to the same results as the chosen $50\times50$ grain. This is for the initial stage where the number density is low and the largest effect would be expected.

The final ice abundances predicted by the KMC simulations are shown in Figures ~\ref{figure-KMC-results-T14ALSO} and \ref{figure-KMC-results-T12ALSO}. We note that due to the long computation times not all models of NIR38 reached the final collapse times and the results shown correspond to slightly shorter collapse times (at least 90\% of the final collapse time; see caption in Figure ~\ref{figure-KMC-results-T14ALSO}). See section \ref{sec:ratevskmc} for a discussion of the limited applicability of the KMC models. 

As shown in Figures ~\ref{figure-KMC-results-T14ALSO} and \ref{figure-KMC-results-T12ALSO}, the absolute column densities of H$_2$O ice tend to be underproduced for all models and towards both sources. While for the shortest collapse time models (10$^4$ yrs) there is a clear mismatch, for the intermediate and longer collapse times ($\geq$10$^5$ yrs), the \ce{H2O} abundance is still within a factor 10 of the observed column density. In terms of ice composition, an increasing trend in abundance is observed for CO, CO$_2$ and CH$_3$OH with increasing density and collapse time, where the models with collapse times $\geq$10$^5$ yrs match better the observations with the exception of \ce{CH4}.  
The ice fractional abundances with respect to H$_2$O, however, remain fairly similar for most species for a given collapse time independently of the density considered in the model. Notable differences are seen for \ce{CO} and \ce{CO2}, where we observe an increasing trend with density, although with the largest collapse time ($t_\textrm{col}$=10$^6$ yr) the fractional abundances level off. Hence, if enough time is given to the ices to grow, these species converge to some final fractional value with respect to \ce{H2O}, which is reached faster for higher densities. 

Comparing \ce{CO2} to \ce{CO}, we find that their final column density ratio is relatively close to 1 (i.e. \ce{CO2}/\ce{CO}$\sim$0.3-3.7). 
Both species tend to have slightly lower fractional abundance than observed for shorter collapse times and lower densities, while the opposite is true for longer collapse times and higher densities. 
For methanol, some differences are found between the simulations for J110621 (Fig.~\ref{figure-KMC-results-T12ALSO}) and for NIR38 (Fig.~\ref{figure-KMC-results-T14ALSO}). For the higher dust temperature models ($T_{\rm dust}=14$~K), the methanol fraction is well reproduced, except for the longest collapse times for which it is overproduced. For the lower dust temperature simulations ($T_{\rm dust}=12$~K), methanol is always slightly overproduced, regardless of collapse times and densities. This might be due to the limited network that misses some reactions leading to more complex molecules such as ethanol. We find that our main formation route for methanol is through successive hydrogenation of \ce{CO} since there is not enough atomic carbon or \ce{CH4} to lead to efficient formation of \ce{CH3OH} from these reactants (Section \ref{sec:ch3oh}). Therefore, the \ce{C + H2O} $\rightarrow$ \ce{H2CO} pathway towards \ce{CH3OH} plays only a minor role.

Unlike the other main ice species, the relative abundances of \ce{NH3} show rather homogeneous results for the different physical conditions with only a minor dependence on dust temperature, collapse time, and density. For \ce{CH4}, however, this species shows larger variations. It is heavily underproduced when compared to the observed value for both sources and it does not obey the upward trend in fractional abundance with collapse time. We find that the formation of \ce{CH4} is hindered by the lack of available atomic carbon on the grain, since elemental carbon is mainly locked up in CO in the gas phase (see Section \ref{sec:phase0}). As reported in Figure \ref{figure-KMC-collapse}, the gas-phase abundances of CO and atomic C show $\ce{CO}/\ce{C} \gg 1$ for all models during the entire duration of collapse. Note that, in fact, this ratio is already rather high at the start of collapse ($\sim 50$), as \ce{CO} is efficiently produced in the gas during the phase 0 translucent cloud stage.

\subsection{Best fit models}
\label{sec:best}

\begin{figure}
\centering
\includegraphics[angle=0,width=1.1\linewidth]{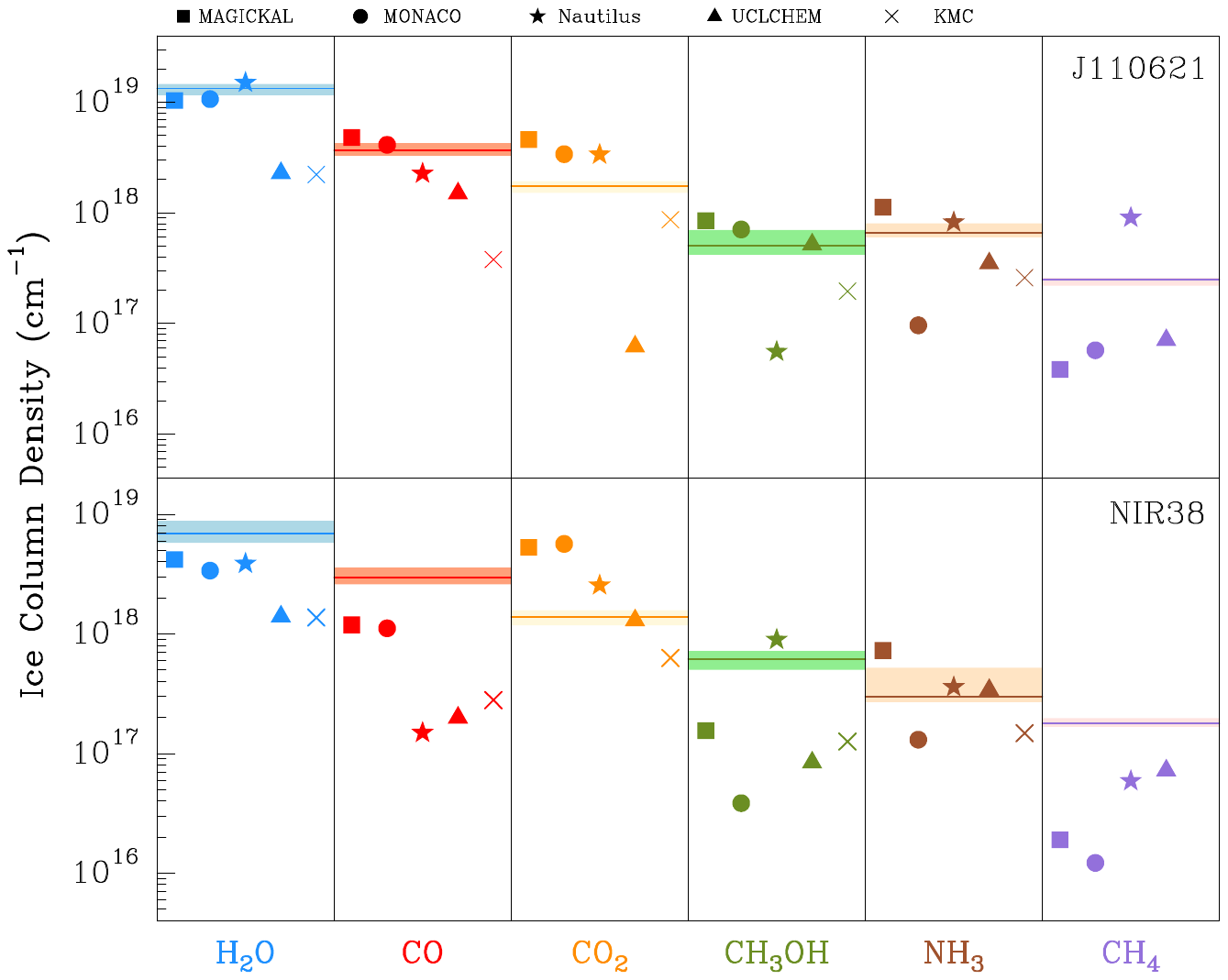}
\caption{Comparison between the ice column densities predicted by all codes for the best fit models, and with the Ice Age observed values. Different colours indicate different molecular species and different symbols label the distinct astrochemical codes used in this work (see legend in the upper part of the Figure). Horizontal lines show the observed ice column densities reported in Table$\,$\ref{tab:observedices} and the shaded areas indicate their associated errors. \label{comparison-best}}
\end{figure}

To identify the model that best matches the Ice Age observations, we calculate the median absolute error (MAE) between the model predictions and the observed column densities in two steps as described in Appendix \ref{app:errors}. This method is also called {\it mean distance of disagreement}, as introduced in \citet{wakelam2006}. In the initial step, we identify a sub-sample of best models characterised by \ce{H2O} column densities consistent with the Ice Age observations within a factor of a few. The models that are favoured are those with longer collapse timescales and higher densities because the efficiency of \ce{H2O} ice accumulation is enhanced and photo-dissociation, a primary mechanism for \ce{H2O} ice destruction in low-density environments, is mitigated.  
In a second step, we obtain the MAE between the predicted and the observed ice abundance ratios for the rest of species with respect to H$_2$O and on logarithmic scale, in order to select the model (or models) with the lowest MAE value (see Appendix \ref{app:errors}). Table \ref{tab:mae} summarises the results of this analysis, and Figure$\,$\ref{comparison-best} presents the comparison of the predicted ice column densities between all codes and with the Ice Age observations.

\begin{table*}
\setlength{\tabcolsep}{10pt} 
\caption{MAE results for the models for NIR38 and J110621.}\label{tab:mae} 
\centering
\renewcommand*{\arraystretch}{1.2}
\begin{tabular}{lcccccc}  
\hline    
& \multicolumn{3}{c}{NIR38} & \multicolumn{3}{c}{J110621} \\
& \multicolumn{3}{c}{($A_{\rm V}$ = 60 mag, $T_{\rm dust}$=14 K)} & \multicolumn{3}{c}{($A_{\rm V}$ = 95 mag, $T_{\rm dust}$=12 K)} \\ \cline{2-7}
Chemical Code & $n_{\rm H}$ ($\rm cm^{-3}$) &  $t_{\rm col}$ (yr) & MAE & $n_{\rm H}$ ($\rm cm^{-3}$) &  $t_{\rm col}$ (yr) & MAE \\ \hline         
MAGICKAL &	2$\times$10$^6$ & 10$^6$ & $0.43_{-0.09}^{+0.11}$ & 2$\times$10$^6$ & 10$^6$ & $0.56_{-0.09}^{+0.10}$ \\
MONACO & 2$\times$10$^6$ & 10$^6$ & $0.63_{-0.10}^{+0.15}$ & 2$\times$10$^6$ & 10$^6$ & $0.45_{-0.08}^{+0.10}$ \\
Nautilus	& 2$\times$10$^4$ & 10$^6$ &	$0.54_{-0.09}^{+0.12}$ & 2$\times$10$^6$ & 10$^6$ & $0.46_{-0.09}^{+0.09}$ \\
UCLCHEM &	2$\times$10$^5$ & 10$^5$ &	$0.48_{-0.14}^{+0.17}$ &  2$\times$10$^5$ & 10$^5$ & $0.62_{-0.12}^{+0.11}$ \\
KMC &	2$\times$10$^5$ & 10$^5$ &	$0.80_{-0.11}^{+0.14}$ & 	2$\times$10$^5$ & 10$^5$ &	$0.76_{-0.08}^{+0.09}$ \\ \hline
\end{tabular}
\tablefoot{The errors associated with the MAE results correspond to the 3$\sigma$ level.}
\renewcommand*{\arraystretch}{1.0}
\end{table*}

For MAGICKAL, the code generally shows best agreement with observations for higher densities and longer collapse timescales. For NIR38 ($A_\textrm{V}$=60 mag, $T_\textrm{dust}$=14 K), the most suitable model conditions are $n_\textrm{H}$=2$\times$10$^{6}$cm$^{-3}$, $t_\textrm{col}=10^6$ yr, with logarithmic MAE value for the ice abundances with respect to \ce{H2O} of $0.43_{-0.09}^{+0.11}$. For J110621 ($A_\textrm{V}=95$ mag, $T_\textrm{dust}=12$K), the observed ice abundance ratios are best explained by the same model conditions, with a corresponding MAE value of $0.56_{-0.09}^{+0.10}$.

Similarly, for MONACO, the best fitting models for both NIR38 and J110621 are the ones with $n_{\rm H} = 2 \times 10^6\, {\rm cm}^{-3}$ and $t_\textrm{col} = 10^6$ yr. The MAE values are $0.63_{-0.10}^{+0.15}$ and $0.45_{-0.08}^{+0.10}$, respectively.

For Nautilus, the best fitting models for NIR38 and J110621 are respectively ($n_{\rm H} = 2 \times 10^4\, {\rm cm}^{-3}$, $t_\textrm{col}$=10$^6$ yr) and  ($n_{\rm H} = 2 \times 10^6\, {\rm cm}^{-3}$, $t_\textrm{col}$=10$^6$ yr).
For NIR38, the model underproduces the \ce{H2O} column density by a factor of $\sim\,$ 1.5--2, with an associated MAE of $0.54_{-0.09}^{+0.12}$. The second best model is the one with the highest density and longest collapse phase ($n_{\rm H} = 2 \times 10^6\, {\rm cm}^{-3}$, $t_\textrm{col}$=10$^6$ yr) with a MAE of $0.95_{-0.10}^{+0.12}$. Although the latter gives in general closer values to the observed \ce{H$_2$O} column density and ice abundance ratios of CO, CO$_2$, CH$_3$OH and NH$_3$, it fails to produce any CH$_4$ and this is the reason for obtaining a worse MAE for this model. 
The observations toward J110621 are better reproduced by the models with an overproduction of \ce{H2O} only by a factor of $\sim\,$ 1.2--1.4 between the modelled and the observed value. The associated MAE is $0.46_{-0.09}^{+0.09}$. Slower collapse overall reproduces more closely the observed column densities.

For \textsc{Uclchem}, the two best models are the ones with intermediate collapse times ($t_{\rm col} = 10^5\, {\rm yr}$) and intermediate or high final densities ($n_{\rm H} = 2 \times 10^5,\, 2 \times 10^6\, {\rm cm}^{-3}$). Between these two models, we cannot favour one over the other, as both the errors in column density and the logarithmic MAE for the abundance ratios with respect to \ce{H2O} show similar values and range between $0.48_{-0.14}^{+0.17}$ and $0.68_{-0.12}^{+0.12}$.

Similarly to \textsc{Uclchem}, the intermediate physical conditions yield the best results for the KMC code. That is, models with collapse time $t_\textrm{col}$=10$^5$ yr and final density $n_{\rm H} = 2 \times 10^5$ cm $^{-3}$ are the ones that best fit the observed ice abundances. These models underproduce the \ce{H2O} column density by a factor of $\sim\,$5 and 8 and show a MAE of $0.8_{-0.11}^{+0.14}$ and  $0.76_{-0.08}^{+0.09}$ for the fractional abundances of NIR38 and J110621, respectively. Other models with lower MAE ($\sim\,$0.6--0.7) are found for shorter collapse times, but they are off in terms of absolute \ce{H2O} column densities with an underproduction by factors $\sim\,$20--70. 

From Figure$\,$\ref{comparison-best}, we find that the largest discrepancies between the best-fit models with the observed ice column densities are found for CH$_3$OH and CH$_4$. These species tend to be underproduced in the ice, especially for the models of NIR38. This is related to the majority of atomic C getting locked into CO in the gas-phase early in the simulations (Section \ref{sec:phase0}), which is subsequently converted into CO$_2$ ice upon accretion, preventing the formation of CH$_3$OH and CH$_4$ in the ice.

To summarise, the models that better reproduce the ice chemical composition observed toward NIR38 and J110621 are those with intermediate/high densities and intermediate/slow collapse times; models with fast collapse and low densities are clearly discarded. These results nicely match the densities and time-scales required by grain growth models ($n_{\rm H}$$\geq$10$^5$$\,$cm$^{-3}$ and t$_{\rm col}$$\geq$10$^6$$\,$yr) to obtain molecular cloud grains of 0.9$\,$$\mu$m in size, which is the size inferred from the JWST Ice Age observations \citep[see][]{dartois2024}.

\subsection{Models with T$_{\rm dust}$=10 K}
\label{sec:lowTdust}

In the model runs explored above, we have considered T$_{\rm dust}$=12 K and 14 K as estimated from {\it Herschel}-based dust temperature maps (see Section \ref{sec:parameters}). However, it is well known that these maps are representative of the outer and slightly warmer layers of molecular clouds \citep[see e.g.][]{hocuk2017,Clement2023}. In order to test the influence of the dust temperature on the ice chemical composition toward the field stars NIR38 and J110621, we have repeated the simulations for the best models identified in Table \ref{tab:mae}, but assuming a dust temperature of $T_{\rm dust}$ = 10 K. The results differ significantly for the codes that include only diffusive chemistry (Nautilus and \textsc{Uclchem}) from those that include both diffusive and non-diffusive chemistry (MAGICKAL, MONACO and KMC). 

For Nautilus, the main impact on the results comes from the severe decrease in the production of CO$_2$ by factors of 55 and 65 for NIR38 and J110621, respectively. At the same time, CO is more abundant in the 10 K models (by factors of 3.6 and 1.8, respectively) as a consequence of a slower diffusion of both CO and O on the surface of dust grains\footnote{The diffusion rate of each species is proportional to $\exp(-E_{\rm bind}/kT_{\rm dust})$, where $E_{\rm bind}$ is the binding energy and $k$ is the Boltzmann constant. This implies that the diffusion rate of atomic O, for instance, decreases by 11 orders of magnitude at 10 K with respect to the case at 12 K.}. CH$_3$OH production is thus increased in both models by factors of 2 and 25 for NIR38 and J110621, while CH$_4$ is also increased in the $T$ = 10 K model, for which CH$_4$ production is in competition with methanol. Similarly to Nautilus, the decrease of $T_{\rm dust}$ in the \textsc{Uclchem} simulations induces a drastic reduction in the formation of \ce{CO2} as well as an increase in the formation of \ce{CH3OH}. This behaviour was hinted from the results for J110621 ($A_{\rm V} = 95$, $T$ = 12 K), for which we already observed an underproduction of \ce{CO2} due to the decrease in the efficiency of the reaction \ce{CO + OH} $\rightarrow$ \ce{CO2 + H}. 

In contrast to \textsc{Uclchem} and Nautilus, we do not observe any significant differences when the temperature decreases down to 10~K for the MAGICKAL, MONACO and KMC codes. For the MAGICKAL code, the chemistry hardly changes for dust temperatures $<$12 K because non-diffusive chemistry takes over diffusive chemistry at these low dust temperatures. As an example, \ce{CO2} predominantly forms at T$_{\rm dust}$=14 K via diffusive association of \ce{CO} and OH. However, for lower T$_{\rm dust}$, the reaction rate of diffusive association of CO and OH declines, and non-diffusive mechanisms begin to dominate the formation of this molecular species. Indeed, the formation rate of both mechanisms becomes equivalent at $\sim$ 13~K. Thus, the decrease in \ce{CO2} (by a factor of 2) is not as drastic as observed for \textsc{Uclchem} or Nautilus, because \ce{CO2} can still form through non-diffusive mechanisms at T$_{\rm dust}$ $<$ 12~K. 

In the case of \ce{CH3OH}, a moderate increase in \ce{CH3OH} ice abundance is predicted (by a factor of 2), which is related to the increase of CO ice abundance as a result of its slower conversion into \ce{CO2} at the lower dust temperature. As shown by the KMC simulations, CH$_3$OH is predominantly formed through \ce{CH3O + H -> CH3OH}, but \ce{CH3O + H2CO -> CH3OH + HCO} also plays a role \citep{Simons2020,santos2022}.  Therefore, the dominant mechanisms responsible for ice formation at T$_{\rm dust}$=10 K for the MAGICKAL, MONACO and KMC codes are non-diffusive, which can operate with the same efficiency independently of the dust temperature considered in the model. 

\section{Discussion}
\label{sec:discussion}

\subsection{Depletion of CO onto ice grains} 
\label{sec:depletion}

From the Ice Age observations of \citet{McClure2023}, CO is the second most abundant species in the ices along the lines of sight of NIR38 and J110621. Despite the significant abundance of CO in solid state ($\sim$30-40\% with respect to H$_2$O; see Table \ref{tab:observedices}), it has been proposed that the cloud local gas density toward these field stars may not exceed the limit of n$_{\rm H}$$\sim$10$^5$ cm$^{-3}$ required for the catastrophic depletion of CO \citep{McClure2023}. This idea is supported by the similar ice chemical composition of the NIR38/J110621 field stars to Elias 16 \citep{McClure2023}, whose visual extinction is $A_{\rm V}$$\sim$19 mag. 

By using the results from the best models reported in Table \ref{tab:mae}, we can provide information about the level of CO depletion predicted for the dense cloud regions toward field stars NIR38 and J110621. The depletion factor of CO, $f_{\rm D}$, is calculated as:

\begin{equation}
    f_{\rm D} = \frac{10^{-4}}{\chi(\ch{CO})_{\rm gas}},
\end{equation}

\noindent
where $\chi$(CO)$_{\rm gas}$ is the gas-phase CO abundance predicted by our best models, and where 10$^{-4}$ is the assumed canonical value for the gas-phase abundance of CO, as measured toward nearby star-forming regions \citep[see e.g.][]{frerking1982,fuente2019}, and that should be considered as a lower limit.

In Table \ref{tab:codep}, we report the values of $\chi$(CO)$_{\rm gas}$ at the end of the collapse and the CO depletion factors predicted by each best model. From Table \ref{tab:codep}, we find that for \textsc{Uclchem} and KMC codes, the CO depletion factors in our best models range between $\sim$2-3. This is produced by the intermediate time-scales for the collapse of their best models, and it would be consistent with the idea of CO not being severely depleted toward the lines of sight of NIR38 and J110621 \citep{McClure2023}. Something similar is found for the Nautilus results of NIR38. However, for J110621, Nautilus gives a very large depletion factor (f$_D$$\geq$1400), which is due to the catastrophic depletion of CO in its highest-density, slowest-collapse model.
The same applies to the MAGICKAL and MONACO codes, for which their predicted depletion factors are clearly $\geq$10. 
Maps of CO ice recently obtained with the JWST toward the Chamaeleon I molecular cloud estimate that between 45\%-85\% of CO is depleted onto dust grains, although large uncertainties do exist \citep[][]{smith2025}. Large-scale maps of gas-phase CO and of its $^{13}$C and $^{18}$O isotopologues toward the Chamaeleon I cloud, will be analysed in the future to accurately measure the level of CO depletion toward the lines of sight of NIR38 and J110621.

\begin{table*}
\setlength{\tabcolsep}{10pt} 
\caption{Predicted CO depletion factors for the best models.}\label{tab:codep} 
\centering           
\begin{tabular}{lcccc}  
\hline\hline    
& \multicolumn{2}{c}{NIR38 ($A_{\rm V}$ = 60 mag)} & \multicolumn{2}{c}{J110621 ($A_{\rm V}$ = 95 mag)} \\
Code & $\chi$(CO)$_{\rm gas}\;(10^{-4})$ & $f_{\rm D}$$^{\rm a}$ & $\chi$(CO)$_{\rm gas}\;(10^{-4})$ & $f_{\rm D}$$^{\rm a}$ \\
\hline         
MAGICKAL & 0.04 &	25 & 0.03 & 30 \\
MONACO & 0.0007 & 1360 & 0.001 & 790  \\
Nautilus &  0.3 & $\sim$3 & 0.0007 & $\geq$1400 \\ 
UCLCHEM & 0.52 & $\sim$2 & 0.53 & $\sim$2 \\
KMC & 0.5 & $\sim$2 & 0.6 & 1.7 \\ \hline
\end{tabular}
\tablefoot{(a) The depletion factor, $f_{\rm D}$, is calculated as 10$^{-4}$/$\chi$(CO)$_{\rm gas}$, with $\chi$(CO)$_{\rm gas}$ the abundance of CO in the gas phase predicted by the best model from each code (see Table$\,$\ref{tab:mae}).}
\end{table*}

\subsection{Formation of CH$_3$OH ice} 
\label{sec:ch3oh}


In all the codes used in this modelling effort, we have considered the same formation routes of methanol on the surface of dust grains. These formation routes are: 1) successive CO hydrogenation in the reaction chain CO $\rightarrow$ HCO $\rightarrow$ H$_2$CO $\rightarrow$ CH$_3$O $\rightarrow$ CH$_3$OH \citep{Watanabe2002}; 2) the reaction chain CH$_4$ + OH $\rightarrow$ CH$_3$ + H$_2$O and CH$_3$ + OH $\rightarrow$ CH$_3$OH \citep{Lamberts2017,Qasim18,Lamberts2022}; and 3) the formation of formaldehyde through C + H$_2$O and subsequent hydrogenation processes \citep{Potapov21, molpeceres2021carbon}. 

For MAGICKAL, \ce{CH3OH} is predominantly ($> 99.99\%$) formed via the successive hydrogenation of CO. The formation of methanol via \ce{CH3} + OH is negligible, given that the hopping rate of heavy species such as radicals is too slow at the low T$_{\rm dust}$ of the simulations. The diffusion of heavy radicals is very slow and the chemistry is dominated by atomic hydrogen at this temperature regime~\citep[$<$20 K;][]{garrod08a}. 
Indeed, for T$_{\rm dust}$=10--16~K, all hydrogen atoms landing on the grain will quickly be involved in a hydrogenation or abstraction reaction and will not desorb, because the surface concentration of reactants is large and the hydrogen atom will only need to diffuse a short distance to react. The same applies to the Nautilus and MONACO codes, for which \ce{CH3OH} is only ($> 99.99\%$) formed by hydrogenation of CO on grains. The methanol formation from \ce{CH3 + OH} $\rightarrow$ \ce{CH3OH} is insignificant, regardless the temperature, density and final collapse time. 

For the KMC simulations, we also find that methanol is predominantly formed through the successive hydrogenation of CO. 
Around 30\% of the total methanol present in the ice has been formed through hydrogenation of \ce{CH3O} whereas the radical-neutral reaction \ce{CH3O + H2CO -> CH3OH + HCO} accounts for $\sim\,$10\% only. The remainder 60\% comes predominantly from hydrogenation of \ce{CH2OH}, which should not be considered as a formation reaction since \ce{CH2OH} is exclusively formed through H-abstraction of methanol. The reaction \ce{CH3 + OH} $\rightarrow$ \ce{CH3OH} hardly occurs in the simulations of both sources (only once in 10$^5$ yr) and the contribution of the \ce{C + H2O} reaction to the formation of formaldehyde is very marginal: only $\sim\,$0.2\% of the reaction flux towards \ce{H2CO} is found to be of this type. 

One may think that the low efficiency in the formation of CH$_3$OH ice through the reaction chain starting from CH$_4$ + OH, is due to the underproduction of CH$_4$ in the models (see Section$\,$\ref{sec:results}). However, note that, even when CH$_4$ is efficiently formed, as in the case of the \textsc{Uclchem} code, the \ce{CH3OH} ice formation reaction \ce{CH3 + OH} $\rightarrow$ \ce{CH3OH} contributes $\leq$1.4\% to the total \ce{CH3OH} ice produced for the best models of NIR38 and J110621.

Therefore, all codes show that the most efficient formation mechanism of methanol for the high extinction conditions probed toward the lines of sight of NIR38 and J110621, is the hydrogenation of CO on grain surfaces. 
We note, however, that we do not exclude that the other two mechanisms of methanol formation (namely, the chain of reactions starting from CH$_4$ + OH and C + H$_2$O) are efficient at lower extinctions and densities, but that the dominant mechanism toward the densest regions in the Chamaeleon I cloud is CO hydrogenation. In fact, the JWST observations reveal that a CH$_3$OH:CO mixture alone is not sufficient to reproduce the observed 9.8$\,$$\mu$m band \citep{McClure2023}, which suggests that some mixing with water may be present toward the outer shells of the cloud (see also Section$\,$\ref{sec:other}).

\subsection{Conversion of CO into CO$_2$}
\label{sec:co2}

Multiple laboratory and theoretical works have been carried out to understand the conversion of CO into CO$_2$ in ices. The most studied surface reactions are: (i) CO + O $\rightarrow$ CO$_2$ \citep[][]{Goumans2010,raut2011,ioppolo2013,minissale2013}; (ii) HCO + O $\rightarrow$ CO$_2$ + H \citep{ruffle2001,ioppolo2013}; and (iii) CO + OH $\rightarrow$ CO$_2$ + H \citep{oba2010,ioppolo2011,noble2011,Molpeceres:2023}. While reactions (i) and (iii) have been found to form CO$_2$ in different quantities, reaction (ii) has been proven to be inefficient \citep{ioppolo2013} and hence, it is not included in the astrochemical codes. Therefore, the main routes for CO$_2$ ice formation are reactions (i) and (iii).

A general result for all models that use directly the values of $T_{\rm dust}$ inferred from Herschel data (14 K and 12 K for NIR38 and J110621, respectively; see Section \ref{sec:results}), is that CO$_2$ is overproduced as compared to the CO$_2$ ice abundances measured with the JWST. Note that the abundances in \citet{McClure2023} could be a factor of $\sim$1.45 higher, due to uncertainties in the determination of the CO$_2$ 4.7$\,$$\mu$m band strength\footnote{The band strength used in \citet{McClure2023} was 1.1$\times$10$^{-16}$$\,$cm$\,$molec$^{-1}$ as revisited by \citet{bouilloud2015}, but a value of 7.6$\times$10$^{-17}$$\,$cm$\,$molec$^{-1}$ can be found in earlier works \citep[][]{gerakines1995}.}. The overproduction of CO$_2$ is a well-known effect that was pointed out by e.g. \citet[][]{garrodandpauly11} and, more recently, by \citet{Clement2023}. 
Indeed, the formation of CO$_2$ ice in the ices is strongly dependent on $T_{\rm dust}$ and it is extremely efficient once $T_{\rm dust}$$\geq$12 K. However, note that the reaction mechanisms responsible for this CO$_2$ ice overproduction is different in different codes. 

For the MAGICKAL code, the dominant  mechanism for CO$_2$ ice formation is the reaction CO + OH $\rightarrow$ CO$_2$ + H when $T_{\rm dust}$$>$12 K. At higher $T_{\rm dust}$, CO becomes mobile enough to reach and react with OH radicals before H or H$_2$, efficiently converting CO into CO$_2$ (see Section \ref{sec:results-magickal}). This reaction is also the dominant CO$_2$ ice formation process in the MONACO and 
\textsc{Uclchem} chemical codes. In contrast, for the Nautilus code, CO$_2$ ice is produced via the grain-surface reaction CO + O $\rightarrow$ CO$_2$, after CO and atomic O being adsorbed onto the icy mantles from the gas phase.  It has been proposed that this difference could be due to distinct binding energies of atomic oxygen used in the codes \citep{Clement2023}. However, in our case, the binding energy used in all codes is $\sim$1600-1660 K, which is consistent
with that recently revisited by \citet{He2015} and \citet{Minissale2022}, and with the lack of detection of O$_2$ in the Orion Bar \citep[][]{melnick2012}. Alternatively, the model discrepancies could be produced by differences in the activation energy barrier assumed for the CO + O $\rightarrow$ CO$_2$ reaction. For example, while MAGICKAL uses 1000 K and 1.27 $\AA$ for the height and width of the reaction barrier \citep[chosen to reproduce the tunnelling rate in the quantum chemical calculations of][]{Goumans2010}, Nautilus uses 627 K and 1 $\AA$, respectively, following the experimental results of \citet{minissale2013}. This would explain why this reaction is more efficient in Nautilus than in MAGICKAL. 
Given all uncertainties, it thus remains unclear which of these two reactions (i.e. CO + OH $\rightarrow$ CO$_2$ + H or CO + O $\rightarrow$ CO$_2$) is responsible for the main production of CO$_2$ ice in the dense molecular regions toward the NIR38 and J110621 field stars. 

In order to investigate this problem in detail, we have used the KMC model where we have included all these different reactions for \ce{CO2} ice formation. First, we considered the reactions where i) \ce{CO2} can form directly from \ce{CO + OH -> CO2 + H}; and ii) \ce{CO2} can form through a minor channel in a two-step fashion \ce{CO + OH -> HOCO} and \ce{HOCO + H -> CO2 + H2}. The model adopts a branching ratio of 0.8 and 0.2 between the two routes. Quantum chemical calculations and laboratory experiments \citep{Arasa:2013, Molpeceres:2023, ishibashi2024} suggest that the latter route should be more important, since HOCO is found to stabilise in the ice rather than to fall apart into \ce{CO2} and H. This latter process is endothermic. 

Additional simulations were performed with an updated network that included  \ce{CO + O -> CO2} \citep[$k=10^{-10}$s$^{-1}$;][]{Goumans2010}, \ce{HCO + O -> CO2 + H} \citep[no barrier;][]{Westenberg1972, Campbell1978} and only the indirect HOCO route with a rate constant of $10^{10}$s$^{-1}$ \citep{Molpeceres:2023}. Interestingly, our results show that all routes contributed roughly equally to the formation of \ce{CO2}, while the rate constants dictate that CO reacting with O should be less favourable by 20 orders of magnitude. The most probable explanation for this is the higher mobility of atomic O on the grain. The species diffuse rapidly in the shallow sites until they reach high binding sites that can act as a ``sink'' where species reside for a long time, long enough to react even for low reaction rate constants. This mechanism is more effective for O than for OH, as a result of its lower diffusion barrier distribution. 
Therefore, the chemical modelling of CO$_2$ on grain surfaces should consider the two formation mechanisms, \ce{CO + O -> CO2} and the indirect HOCO route.

\subsection{Formation of hydrides in the ice: H$_2$O, NH$_3$ and CH$_4$}
\label{sec:hydrides}

As shown in Section \ref{sec:results}, the models tend to underproduce water for all codes. The models that generate the largest amounts of water are the ones with a slow-collapse phase and with high-densities, reproducing the observed water ice abundances within a factor of 10.

Similarly, CH$_4$ also tends to present lower abundances in all model runs as compared to the ones measured by the Ice Age observations. An exception is the \textsc{Uclchem} code, for which the models reproduce within factors of 2-4 the CH$_4$ ice abundances observed toward NIR38 and J110621 (see Figure$\,$\ref{comparison-best}). The formation of CH$_4$ ice is intimately linked to the formation of CO$_2$ ice since both molecular species compete for the conversion of atomic C depleted onto dust grains. This conversion, however, is expected to take place at lower extinctions during the translucent cloud (phase 0) stage \citep[see e.g.][]{garrodandpauly11}. As explained in Section \ref{sec:phase0}, the amount of gas-phase atomic C available for depletion in the translucent cloud (phase 0) stage is roughly one order of magnitude higher in the \textsc{Uclchem} models than for the other codes. For NH$_3$, its formation in the ice occurs predominantly via the hydrogenation of atomic N. The models reproduce fairly well the ice abundances observed with the JWST, which differ by typically factors of a few.

We finally note that lowering the temperature from 12~K/14~K to 10~K in our simulations has, in general, little to no effect on the resulting ice abundances of H$_2$O, CH$_4$ and NH$_3$. This is particularly true for the codes that include non-diffusive chemistry such as MAGICKAL, MONACO and the KMC codes.

\subsection{Oxygen budget in the models}
\label{sec:oxygen}
For our modelling, we have considered an initial abundance of 1.76$\times$10$^{-4}$ for atomic oxygen, as proposed in the $"$EA1$"$ low-metal abundance case of \citet[][see Table$\,$\ref{tab:initial}]{Wakelam08}. We can check whether this abundance is consistent with the oxygen budget locked into H$_2$O$_{\rm ice}$, CO$_{\rm ice}$, and CO$_{\rm 2,ice}$ in the models. Taking into account that our best models provide values close to those measured with the JWST, we use the abundances shown in Table$\,$\ref{tab:observedices} to make this estimate. For NIR38, the oxygen elemental abundance locked into the ice can be calculated from the sum N(H$_2$O$_{\rm ice}$) + N(CO$_{\rm ice}$) + 2$\times$N(CO$_{2,ice}$), which gives $\sim$1.3$\times$10$^{19}$$\,$cm$^{-2}$. Since the column density of hydrogen nuclei toward NIR38 is $\sim$10$^{23}$cm$^{-2}$, this yields an oxygen abundance of 1.34$\times$10$^{-4}$, which corresponds to 76\% of the initial oxygen assumed in the models. However, the "EA1" abundance of 1.76$\times$10$^{-4}$ considers depletion onto dust grains and hence, when compared to the cosmic abundance of oxygen \citep[$\sim$4.9$\times$10$^{-4}$; see][]{asplund2009}, the oxygen abundance locked into ices represents 27\% of the total atomic oxygen. On the other hand, the derived oxygen abundance in the ice toward the J110621 line-of-sight is $\sim$2.1$\times$10$^{-4}$, i.e. higher than the initial atomic oxygen abundance assumed in our models. Therefore, the "EA2" and "EA3" cases proposed by \citet{Wakelam08}, provide a more realistic initial  abundance for elemental oxygen for astrochemical modelling of dense molecular clouds.

\subsection{Importance of the length of the collapse}
\label{sec:collapse}

In our models, we have considered different collapse time lengths to establish the impact of this parameter on the ice chemical composition. As shown in Section \ref{sec:results}, the abundances of major ice species such as H$_2$O, CO or CO$_2$ change little in the fast-collapse models because the collapse time length ($t_{\rm col}$=10$^4$ yrs) is orders of magnitude shorter than the depletion timescale ($t_{\rm freeze-out} \simeq 10^9/n_H$\,yr). Only for the highest-density models with n(H)=10$^6$ cm$^{-3}$, for which the depletion time-scales are similar to the collapse time length, the predicted ice abundances reach values closer to those measured in the Ice Age observations. As discussed in Section \ref{sec:best}, the best agreement with the observations is achieved for models with longer collapse time lengths (i.e. with time-scales $t_{\rm col}$=10$^5$ yrs and $t_{\rm col}$=10$^6$ yrs; see Table \ref{tab:mae}), although note that shorter collapse times favour the formation of species such as CH$_4$ (see Figures \ref{figure-KMC-results-T12ALSO} and \ref{figure-KMC-results-T14ALSO}). The high densities and long timescales for ice formation are broadly consistent with the $n_H$=10$^5$ cm$^{-3}$ and t=10$^6$ years required by grain growth models for molecular clouds \citep[][]{ormel2009} in order to reproduce the 0.9 $\mu$m grain sizes found for NIR38 and J110621 by \citet{dartois2024}. The convergence of models for two different physical and chemical phenomena adds weight to these results.

\subsection{Uncertainties in the determination of the extinction toward NIR38 and J110621}
\label{sec:Av}



The visual extinction values first reported in \citet{McClure2023} toward the lines-of-sight of NIR38 and J110621 ($A_{\rm V}$=60 mag and $A_{\rm V}$=95 mag, respectively), were pre-flight estimates.  
A more recent analysis of the NIRSpec spectra obtained with the JWST toward these two field stars, has however yielded A$_K$ values of 4.1 for NIR38 and 5.6 mag for J110621 \citep{dartois2024}. If we assume $A_{\rm V}$/A$_K$ = 8.4, these values correspond to visual extinctions of $A_{\rm V}$=34 mag and $A_{\rm V}$=47 mag, respectively. These values are different from those given by \citet{McClure2023}, because these authors used the pre-launch extinctions determined photometrically under the assumption that both background stars were both G or K giants. Instead, \citet{dartois2024} fitted the photospheric absorption features to determine that both stars were K7 dwarfs.

The new $A_{\rm V}$ values are therefore factors of $\sim$2 lower than those reported in \citet{McClure2023} and than those used in our chemical modelling simulations (see Section \ref{sec:models} and Table \ref{tab:gridofmodels}), which explains the similar ice compositions of NIR38/J110621 and Elias 16. However, we note that we do no expect significant differences for these lower visual extinctions because UV photo-chemistry (the only one that could be affected by the lower extinction values) is inefficient at $A_{\rm V}$$>$10 mag. The only impact is expected for the predicted column densities of water ice reported in Figures \ref{magickal-results-Av60}-\ref{figure-KMC-results-T12ALSO}, whose values would be lower by roughly a factor of $\sim$2. The ice column density ratios would not suffer any change because they are calculated directly from the ice abundances predicted by the models. The predicted CO depletion factors would not be affected either by the change in visual extinction for NIR38 and J110621, because f$_D$ are inferred from the CO gas-phase abundance given by the models (see Section \ref{sec:depletion}).


\subsection{Other possible effects not considered in the models: Grain growth, cloud dynamics and line-of-sight effects}
\label{sec:other}

The aggregation and clustering of icy dust grains during the collapse of a cloud will modify their size distribution, as explored numerically in e.g. \citet{Weingartner2001,ormel2011dust,Silsbee2020,Marchand2023,Lebreuilly2023}. For the two background sources behind Chamaeleon I studied here, NIR38 and J110621, it has been shown by simulating the observed absorption spectra that the largest grains in the distribution have increased in radius by a factor of about four \citep{dartois2024} with respect to the largest grains in the expected diffuse ISM distribution  \citep{Mathis1977}, giving rise to clear scattering features. 
Enhanced scattering observed more widely across this cloud with Spitzer photometric filters has also been attributed to grain growth \citep{Lefevre2014}. 
Grain growth is accompanied by a lowering of the visual extinction per dust mass (see Section \ref{sec:Av}) and is expected to lower the peak dust temperatures and modify the radiative and sublimation-induced cooling efficiencies of dust grains \citep{HerbstCuppen2006}.  
In addition, the surface to volume ratio might be affected and thus influence, possibly diminish, the available total surface for chemistry and the surface available for depletion from the gas phase. Clues as to the local ice structure and morphology were given by the observation of the ``dangling OH'' absorption features of H$_2$O ice along both lines of sight \citep{Noble2024}. These features, while not providing a direct measure of porosity/compaction of the ice, trace H$_2$O molecules not fully bound to surrounding H$_2$O molecules in the ice, whether due to interaction with other molecules such as CO, CO$_2$, or due to a non-bound OH in the bulk or at the ice surface. The surface to volume parameter will depend on the degree of compaction of the ice during the aggregation process. 
Modification of the grain surface morphology due to energy dissipation following grain collision or cosmic ray hits can also impact the chemistry that occurs in and on the grain, as the availability and nature of binding sites changes. All of these physical modifications to the grain will, in turn, modify binding energy distributions at the surface and in the bulk, and thus activation barriers to diffusion and adsorption/desorption \citep{Noble2012,Cuppen2013,Karssemeijer2014,Nguyen2018,Minissale2022},  impacting the availability of reactants for, and the relative rates of, the various chemical processes discussed above \citep{HerbstCuppen2006,Penteado_ea17,Iqbal2018}. 

Another parameter that could affect the results of our simulations is the history of the evolution of the molecular cloud. In the simulations presented in this paper, we have considered a simple evolution of the physical properties of the cloud (temperature and density) during its collapse phase \citep[we have assumed free-fall collapse; see Section \ref{sec:parameters} and][]{brown88,rawlings1992}. However, by doing so, we have only considered a small portion of the possible evolutionary pathways that a parcel of gas in the Chamaeleon I cloud could undergo. By using hydrodynamical simulations, \citet{Clement2023} have shown that the passage from the diffuse to dense cloud phase in these simulations, was key to producing CO$_2$ very early in the cloud's evolution when the dust temperature is larger than 12~K, but also to generate enough hydrides before atoms are fully converted into other molecules in the gas-phase. In addition, note that, for the sake of simplicity, we have kept the dust and gas temperature equal in our simulations, and have varied them only between 14 and 10~K. As explored by \citet{Clement2023}, the decoupling between the gas and dust temperatures may lead to different ice chemical compositions. 

Finally, the model results compared to the JWST observations are those at the end of the phase 1 collapse stage, which is characterised by constant physical conditions. In reality, JWST observations toward the two field stars NIR38 and J110621 are probing all material along the two lines of sight from the outer edges to the central zone of the molecular cloud. Therefore, line-of-sight effects due to the inhomogeneous structure of the molecular cloud are not taken into account by the modelling. This could be one of the reasons behind the general difficulty of the models to reproduce the observed CH$_3$OH and CH$_4$ ice composition simultaneously, as these two observed ice components may probe different locations along the line of sight. For example, CH$_4$-rich ice could be present in the outer shell of the cloud, where the C to CO conversion is not complete (so that C atoms can be efficiently hydrogenated once accreted onto dust grains). In contrast, CH$_3$OH ice should copiously form toward the central colder and denser regions of the cloud, where most atomic C is locked into CO and CO can efficiently freeze out onto the surface of dust grains to be transformed into CH$_3$OH via successive hydrogenation.
However, note that visual extinctions A$_{\rm V}$$\leq$2$\,$mag are required for carbon to remain in its atomic form in the gas phase. Indeed, as shown in Figure$\,$\ref{figure-staticphase}, even in the translucent cloud regime with visual extinctions A$_{\rm V}$=2$\,$mag, atomic C is almost completely converted into gas-phase CO. This implies that the region where atomic C could avoid conversion into CO extends at most A$_{\rm V}$=2$\,$mag into the cloud from the surface.
Considering that every cloud has two surfaces -- the one on the observer's side and the second one on the far side -- the surface layer would extend for 4$\,$mag in total, which represents $\leq$12\% of the NIR38 line of sight, assuming that its extinction is at least A$_{\rm v}$=34$\,$mag (as mentioned in Section$\,$\ref{sec:Av}).
Detailed physical and chemical modelling of the cloud is needed to reconstruct such line-of-sight effects, and this requires a dedicated observational effort, including dust continuum and spectral line emission mapping, which is undergoing. 


\subsection{Differences between diffusive versus non-diffusive chemistry}
\label{sec:diffusive}

In this work, we have used astrochemical codes that consider diffusive chemistry only (e.g. Nautilus, \textsc{Uclchem}) and codes that include both diffusive and non-diffusive chemistry (MAGICKAL, MONACO, KMC). The largest differences between these codes are mainly found for models with very low dust temperatures of T$_{\rm dust}$=10 K. Indeed, in codes with diffusive-chemistry-only, diffusion of grain-surface species becomes activated at higher T$_{\rm dust}$. This is the reason why the production of species such as CO$_2$ is largely decreased, in favour of CH$_3$OH and CH$_4$. In contrast, for codes that include diffusive and non-diffusive chemistry, the formation of new molecular species at T$_{\rm dust}$=10 K occurs exclusively by non-diffusive processes (i.e. {\it in-situ} on the grain surface upon the formation of the reactants) and therefore, it does not require the species to diffuse across the surface. This implies that non-diffusive chemistry is efficient at T$_{\rm dust}$=10 K. As shown in Section \ref{sec:lowTdust}, no differences are found between models with T$_{\rm dust}$=10, 12 and 14 K for codes with non-diffusive chemistry, which indicates that the formation of H$_2$O, CO, CO$_2$, CH$_3$OH, CH$_4$ and NH$_3$ in these models occurs mainly non-diffusively at these T$_{\rm dust}$. This could explain the rather constant ice chemical composition observed not only in Chamaeleon I but also in other dense molecular clouds \citep[as e.g. toward the Elias 16 background star;][]{Knez2005}, as found by \citet{McClure2023}.

\subsection{Differences between rate equation codes versus stochastic KMC models}
\label{sec:ratevskmc}

Rate equation codes are macroscopic models in which every different chemical reaction results in a term in the rate equation that combines the meeting of the reactants and the subsequent reaction. Whether the reactants meet by diffusion or through some non-diffusive mechanism should be explicitly included in the rate equations, thereby requiring an estimation of the coupling between the different processes and the average outcome.

In contrast, the lattice KMC model is microscopic in the sense that it simulates every move of each individual species on top of the grain surface, monitoring their exact location. In this model, the effect of diffusion on reactions is observed as an outcome of the simulation of each individual move of a species. This can pose an advantage, especially when the coupling between processes is difficult to model or when assumptions for the rate equations do not hold throughout the entirety of evolution of the system. 
Another advantage is that the surface structure of the grain can be taken into account. The rates of various processes (e.g. diffusion, reaction and desorption) is made site-specific and so the effect of surface morphology on the surface chemistry can be modelled in detail.

On the other hand, the microscopic approach of the lattice KMC grain model comes with a cost in terms of practical applicability. Certainly, when compared to macroscopic rate equation models, the computation time is very long. Even so much so, that for the highest final densities and longest collapse times of the grid studied in this work, the computation time becomes inapplicable (estimated time is on the order of months). Therefore, for some combinations of parameters, we have not been able to include results for the entire collapse time duration (instead at least $94\%$ of collapse time). Moreover, as a single KMC run simulates \emph{one possible} (stochastic) evolution of the system (grain), ideally one would like to perform multiple runs and average over the results. Evidently, this is currently unfeasible for the densities and collapse times studied here.

Long computation times also place restrictions on the resolution that we can use with our KMC model. As is obvious from the \ce{CH4} abundance, which we found to fluctuate between 1 and 0 on the grain, we would like to use larger lattice size $N$. However the simulation time scales with N$^2$log(N), making higher resolution simulations unattainable for the present effort. Conversely, if one would like to reduce the computation time, the discrete fluctuations in abundance for some species shows that we are already at the limit of being able to resolve one particle in our system.

 
\section{Conclusions}

\label{sec:conclusions}

In this work, we have used state-of-the-art astrochemical codes (MAGICKAL, MONACO, Nautilus, \textsc{Uclchem} and KMC) to model the ice abundances measured with the JWST within the Ice Age ERS program toward the highly-extinguished background stars NIR38 and J110621 in the line of sight of the Chamaeleon I dense molecular cloud \citep[see][]{McClure2023}. To do this, we have considered a small grid of physical parameters covering gas densities $n_{\rm H}$=2$\times$10$^4$, 2$\times$10$^5$ and 2$\times$10$^6$ cm$^{-3}$, dust temperatures of T$_{\rm dust}$=10, 12 and 14 K, and cloud collapse time-scales of t$_{\rm col}$=10$^4$, 10$^5$ and 10$^6$ yr. 
The comparison of the model predictions with the observed JWST ice abundances toward NIR38 and J110621 allow us to understand the chemical processes behind the ice formation of CH$_3$OH (and hence, of CO and CO$_2$, which are tightly related to the ice chemistry of CH$_3$OH) and of hydride species such as H$_2$O, CH$_4$ and NH$_3$. The main conclusions of our work are listed below:

\begin{enumerate}
\item 
The best match between the predicted and the observed ice abundances is found for the higher-density and slower collapse models with $n_{\rm H}$$\geq$2$\times$10$^5$ cm$^{-3}$ and t$_{\rm col}$$\geq$10$^5$ yr. Low-density ($n_{\rm H}$=2$\times$10$^4$ cm$^{-3}$), fast models (t$_{\rm col}$=10$^4$ yr) are ruled out because their collapse time-scales are shorter that the depletion time-scales preventing the efficient growth of the ices. These results nicely match the densities and time-scales required by grain growth models to grow molecular cloud dust grains to sizes of 0.9$\,$$\mu$m as measured toward the Chamaeleon I cloud.

\item 
The models provide mixed values for the CO depletion factor, which range from f$_D$$\sim$2-3 to f$_D$$\geq$25. Large-scale maps of gas-phase CO are therefore needed to constrain the level of CO depletion onto dust grains.

\item 
All our models find that CH$_3$OH ice formation toward the highly-extinguished lines-of-sight of NIR38 and J110621 in the Chamaeleon I cloud, predominantly occurs via the successive hydrogenation of CO. The contribution of other proposed CH$_3$OH formation mechanisms, such as the reaction chains starting from CH$_4$ + OH or from C + H$_2$CO, is minor for the high-extinction conditions probed toward the NIR38 and J110621 field stars.  

\item 
Our modelling effort shows that CO$_2$ ice abundance tends to be overpredicted in all models, but significant discrepancies are found across codes regarding the dominant reaction for CO$_2$ ice formation (i.e. reaction CO + OH forming an HOCO intermediate versus reaction CO + O $\rightarrow$ CO$_2$). Our KMC simulations, however, reveal that both reactions should be considered in any ice chemical modelling given that they contribute equally to the formation of CO$_2$ ice as a result of the higher mobility of atomic O on the grain as compared to OH. 

\item 
The formation of H$_2$O ice tends to be underpredicted in all models, although the predicted values lie within a factor of 10 with respect to the JWST observations. CH$_4$ is largely underproduced for all codes except for \textsc{Uclchem}, which shows a higher gas-phase abundance of atomic C during the translucent cloud phase due to more efficient non-thermal desorption. The models reproduce fairly well the observed ice abundance of NH$_3$. 

\item 
Models with dust temperatures T$_{\rm dust}$=10 K show significant changes in the results for codes with diffusive chemistry only as compared to the results with T$_{\rm dust}$=12/14 K (i.e. Nautilus and \textsc{Uclchem}). However, no differences are found for codes that consider non-diffussive chemistry. This is due to the fact that non-diffusive mechanisms take over at lower T$_{\rm dust}$, compensating for the less efficient diffusiveness of atomic species and radicals at these low dust temperatures. 

\end{enumerate}

As discussed in Section \ref{sec:other}, our chemical modelling does not consider the effects of grain growth and/or ice morphology \citep[][]{dartois2024,Noble2024}, cloud dynamics \citep{Clement2023} or the presence of multiple ice components along the line-of-sight. Future chemical modelling will have to be carried out to assess the importance of all these effects in detail. Furthermore, spatial information on the gas-phase abundances of these major ice compounds will be obtained in the near future to complement the JWST Ice Age observations, which will help us to constrain key parameters in the modelling of the chemistry of ices (such as the depletion of CO, f$_D$) toward the Chamaeleon I molecular cloud.  

\begin{acknowledgements}
I.J.-S. would like to thank all the modellers involved in this project, who made it an enjoyable and stimulating experience. 
We also would like to thank Sergio Ioppolo and Giulia Perotti for insightful discussions and useful comments.
I.J.-S. acknowledges funding from the European Research Council (ERC) under European Union's Horizon Europe research and innovation program OPENS (101125858). Views and opinions expressed are however those of the author(s) only and do not necessarily reflect those of the European Union or the European Research Council Executive Agency. Neither the European Union nor the granting authority can be held responsible for them. I.J.-S. and A.M. also acknowledge funding from grant PID2022-136814NB-I00, and grant PRE2019-091471, funded by the Spanish Ministry of Science, Innovation and Universities / State Agency of Research, MCIN/AEI/10.13039/501100011033, and by $``$ERDF/EU, A way of making Europe$"$ and $``$ESF, Investing in your future$''$. V.W., A.T., J.A.N. and E.D acknowledge support from the French program $``$Physique et Chimie du Milieu
Interstellaire$''$ (PCMI) of the CNRS/INSU with the INC/INP cofunded by the CEA and CNES. MJ was supported by the NASA Planetary Science Division Internal Scientist Funding Program through the Fundamental Laboratory Research work package (FLaRe). R.T.G. thanks the National Science Foundation for funding through the Astronomy \& Astrophysics program (grant number 2206516). 
The work by A.I.V. and K.B. has been supported via the project number FEUZ-2020-0038. M.N.D. acknowledges the Holcim Foundation Stipend. S.V. acknowledges funding from the European Research Council (ERC) under the European Union's Horizon 2020 research and innovation program MOPPEX 833460. Part of this research was carried out at the Jet Propulsion Laboratory, California Institute of Technology, under a contract with the National Aeronautics and Space Administration (80NM0018D0004). D.C.L. acknowledges financial support from the National Aeronautics and Space Administration (NASA) Astrophysics Data Analysis Program (ADAP). Astrochemistry at the Open University is currently supported by STFC under grant agreement numbers ST/X001164/1, ST/T005424/1 and ST/Z510087/1 and H.J.F. also thanks The Open University for supporting the OU Astrochemistry Group. Astrochemistry in The Netherlands is supported by the NWO Dutch Astrochemistry Network (grant no. ASTRO.JWST.001).
\end{acknowledgements}


\bibliography{aa}{}
\bibliographystyle{aa}




\begin{appendix}

\section{Additional grain-surface chemical reactions included in \textsc{Uclchem} for HOCO and $\ce{CO2}$} 
\label{app:uclchem-hoco}

For the \textsc{Uclchem} calculations, we have added several chemical reactions involved in the grain-surface chemical network of the hydrocarboxyl radical (HOCO) and $\ce{CO2}$. These reactions are listed in Table \ref{table-uclchem-extra-reactions}, and they have been extracted from the works by \cite{Goumans2008}, \cite{Goumans2010}, \cite{Arasa:2013},  \cite{Ruaud2015}, \cite{Qasim2019}, and \cite{Molpeceres:2023}.

\begin{table}[h!]
\caption{Grain-surface reactions included in \textsc{Uclchem} for HOCO and $\ce{CO2}$.}
\label{table-uclchem-extra-reactions}
\centering
\renewcommand*{\arraystretch}{1.2}
\begin{tabular}{cccc}
\hline
\multicolumn{3}{c}{Reaction} & Energy barrier (K) \tabularnewline
\hline
${\rm CO+O~\rightarrow}$ & ${\rm CO_2}$ &  & 2500 \tabularnewline
\rule{0pt}{4ex} 
\multirow{2}{*}{${\rm CO+OH~\rightarrow}$} & ${\rm HOCO}$ & (50\%) & 150\tabularnewline
 & ${\rm H+CO_{2}}$ & (50\%) & 150\tabularnewline
\rule{0pt}{4ex} 
\multirow{2}{*}{${\rm N+HOCO~\rightarrow}$} & ${\rm NH+CO_{2}}$ & (50\%) & 0\tabularnewline
 & ${\rm OH+OCN}$ & (50\%) & 0\tabularnewline
 \rule{0pt}{4ex} 
${\rm O+HOCO~\rightarrow}$ & ${\rm OH+CO_{2}}$ &  & 0\tabularnewline
\rule{0pt}{4ex} 
\multirow{3}{*}{${\rm H+HOCO~\rightarrow}$} & ${\rm H_{2}+CO_{2}}$ & (70\%) & 0\tabularnewline
 & ${\rm CO+H_{2}O}$ & (20\%) & 0\tabularnewline
 & ${\rm HCOOH}$ & (10\%) & 0\tabularnewline
\hline
\end{tabular}
\renewcommand*{\arraystretch}{1.0}
\tablefoot{The third reaction, forming \ce{H} and \ce{CO2}, was already present in the default chemical network of \textsc{Uclchem}, but we updated the energy barrier from 1000 K to 150 K \citep{Ruaud2015}.}
\end{table}

\section{Evolution of the ice and gas phase abundances for each code} 
\label{sec:evolution}

We present here the evolution of the ice abundances obtained during the collapse (phase 1) stage by the MONACO, Nautilus, \textsc{Uclchem} and KMC models for H$_2$O, CO, CO$_2$, CH$_3$OH, CH$_4$ and atomic C. Four representative models are shown corresponding to the cases of a low-density, fast collapse model and of a high-density, slow collapse model for both field stars NIR38 and J110621. In these Figures, we also show the evolution of the gas phase abundances of CO and atomic C. 

The evolution is similar for all codes: the ices undergo a fast growth by the end of the collapse for the high-density, slow collapse models, when the density rapidly increases and the catastrophic depletion of CO takes place (see right panels in these Figures). We note that the evolution of the density follows the free fall equations, meaning that the increase of density is very rapid towards the end of the calculations. As a consequence, for the models with final n$_{\rm H}$ densities of $2\times 10^4$~cm$^{-3}$, the increase from about $2\times 10^3$~cm$^{-3}$ to $2\times 10^4$~cm$^{-3}$ is done very rapidly (regardless the collapse timescales) and thus, the ice chemical composition does not have time to evolve and it reflects the chemistry at very low densities inherited from the beginning of the collapse. Therefore, the ices do not experience significant growth.   

\begin{figure*}
\begin{center}
\includegraphics[width=1\linewidth]{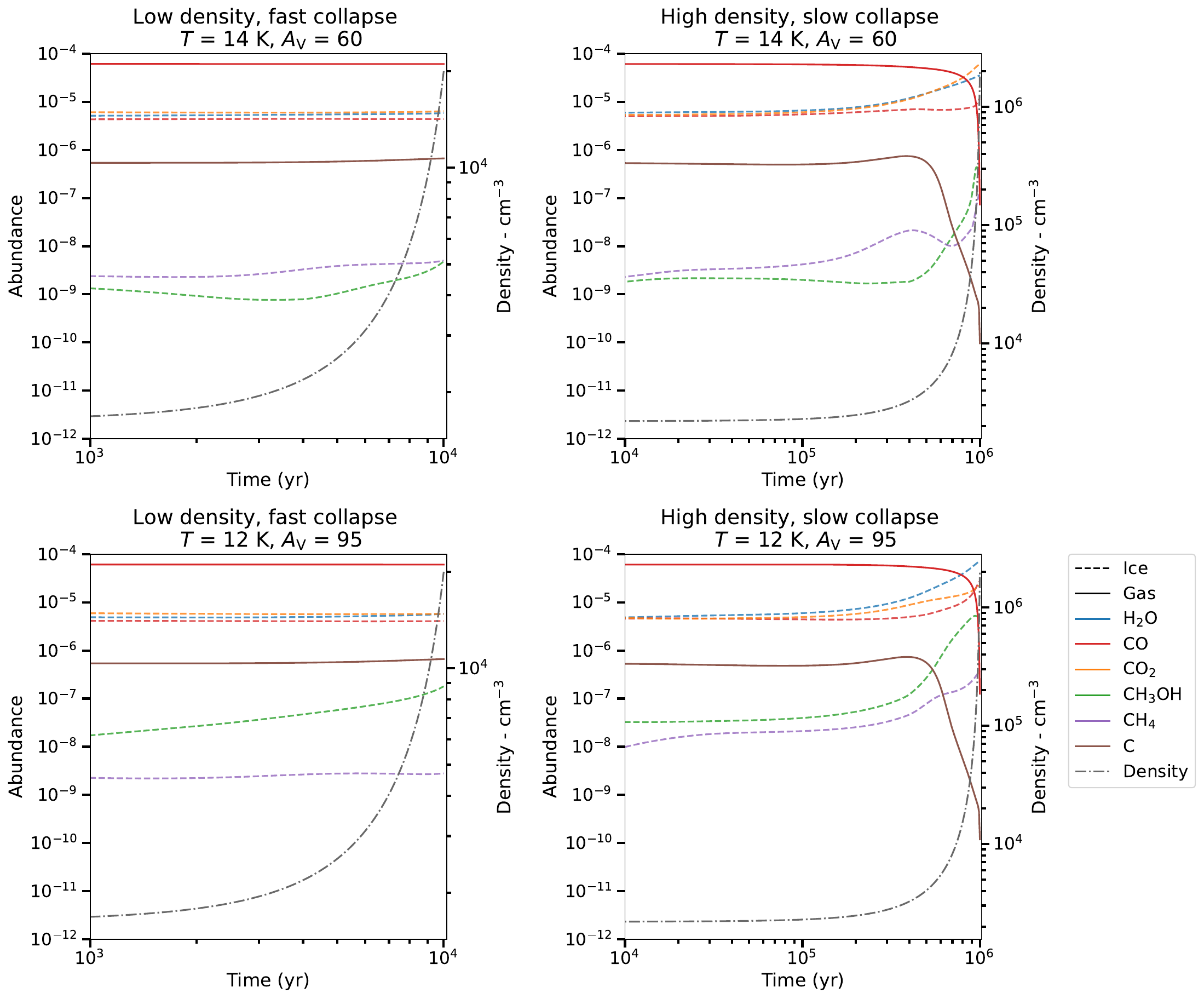}
\caption{Time evolution for most relevant species predicted by some of our MONACO models (the extreme cases) for the collapse phase.}
\label{figure-MONACO-collapse}
\end{center}
\end{figure*}

\begin{figure*}
\begin{center}
\includegraphics[width=1\linewidth]{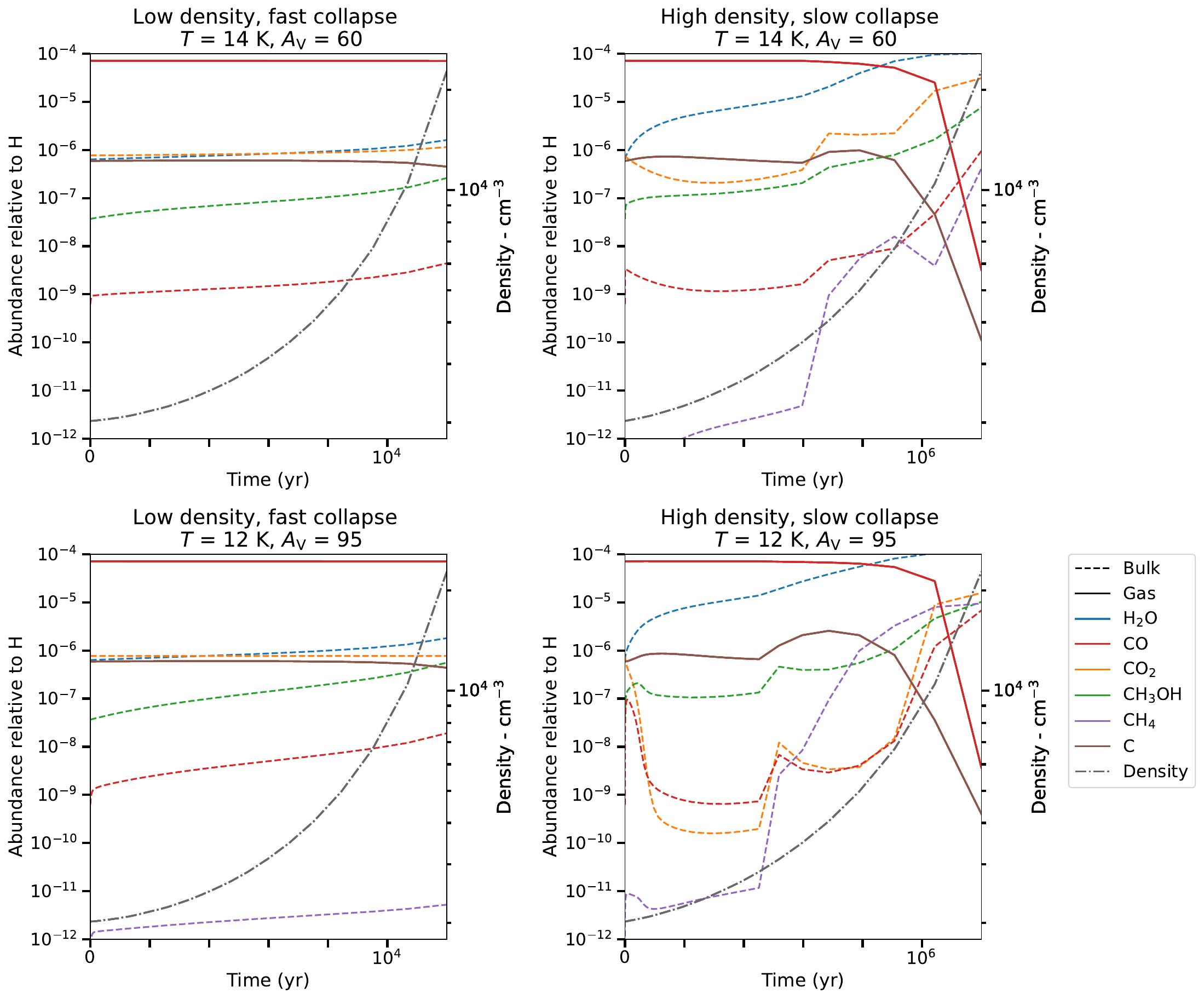}
\caption{Time evolution for most relevant species predicted by some of our \textsc{Nautilus} models (the extreme cases) for the phase 1 collapse phase.}
\label{figure-nautilus-collapse}
\end{center}
\end{figure*}

\begin{figure*}
\begin{center}
\includegraphics[width=1\linewidth]{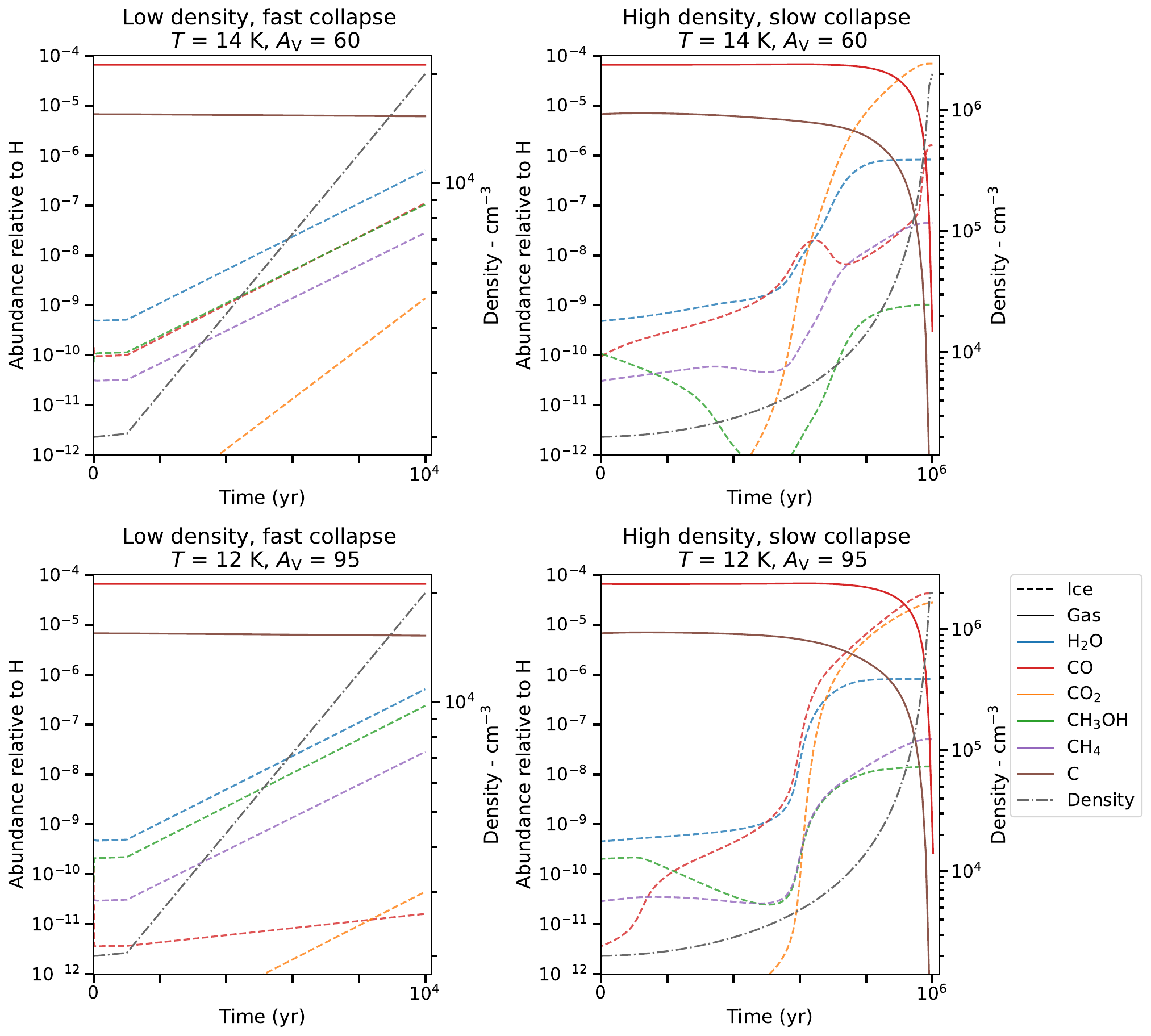}
\caption{Time evolution for most relevant species predicted by some of our \textsc{Uclchem} models (the extreme cases) for the collapse phase.}
\label{figure-uclchem-collapse}
\end{center}
\end{figure*}

\begin{figure*}
\begin{center}
\includegraphics[width=1\linewidth]{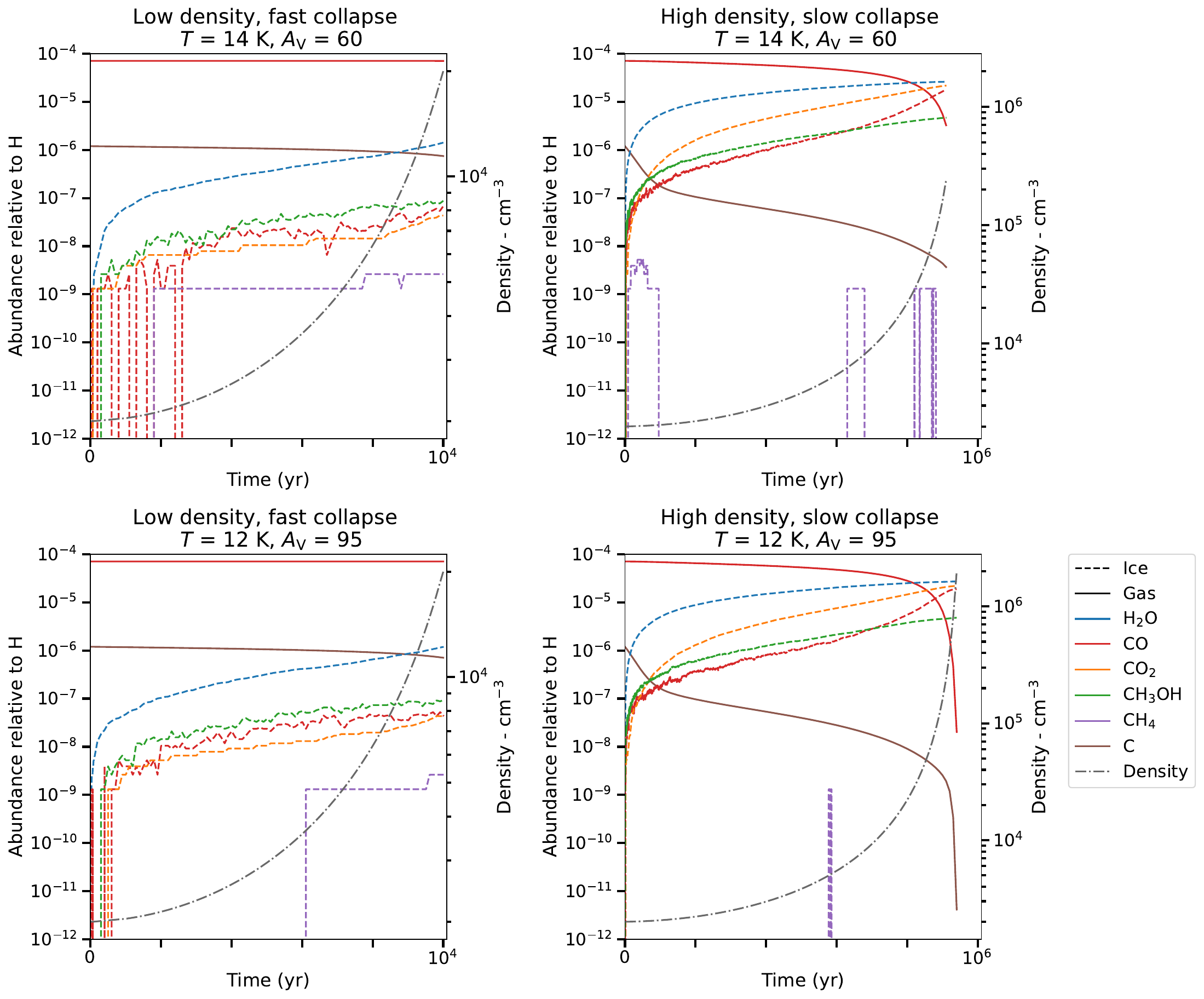}
\caption{Time evolution for most relevant species predicted by some of our \textsc{KMC} models (extremes) for the collapse phase. Note that the heavy fluctuations observed for \ce{CH4} is a finite-size effect, due to the very low concentration numbers generated by the KMC on the grain (CH$_4$ abundance is found to fluctuate between 1 and 0 on our $50\times 50$ lattice). For the high-density slow-collapse model (model 9) towards NIR38 ($T = $ 14 K, $A_\mathrm{V}$ = 60 mag), the results are shown for 95.9 \% of the intended collapse time.}
\label{figure-KMC-collapse}
\end{center}
\end{figure*}

\section{Predicted ice composition and comparison with Ice Age observations} \label{app:histograms}

\begin{figure*}
\centering
\includegraphics[width=1\linewidth]{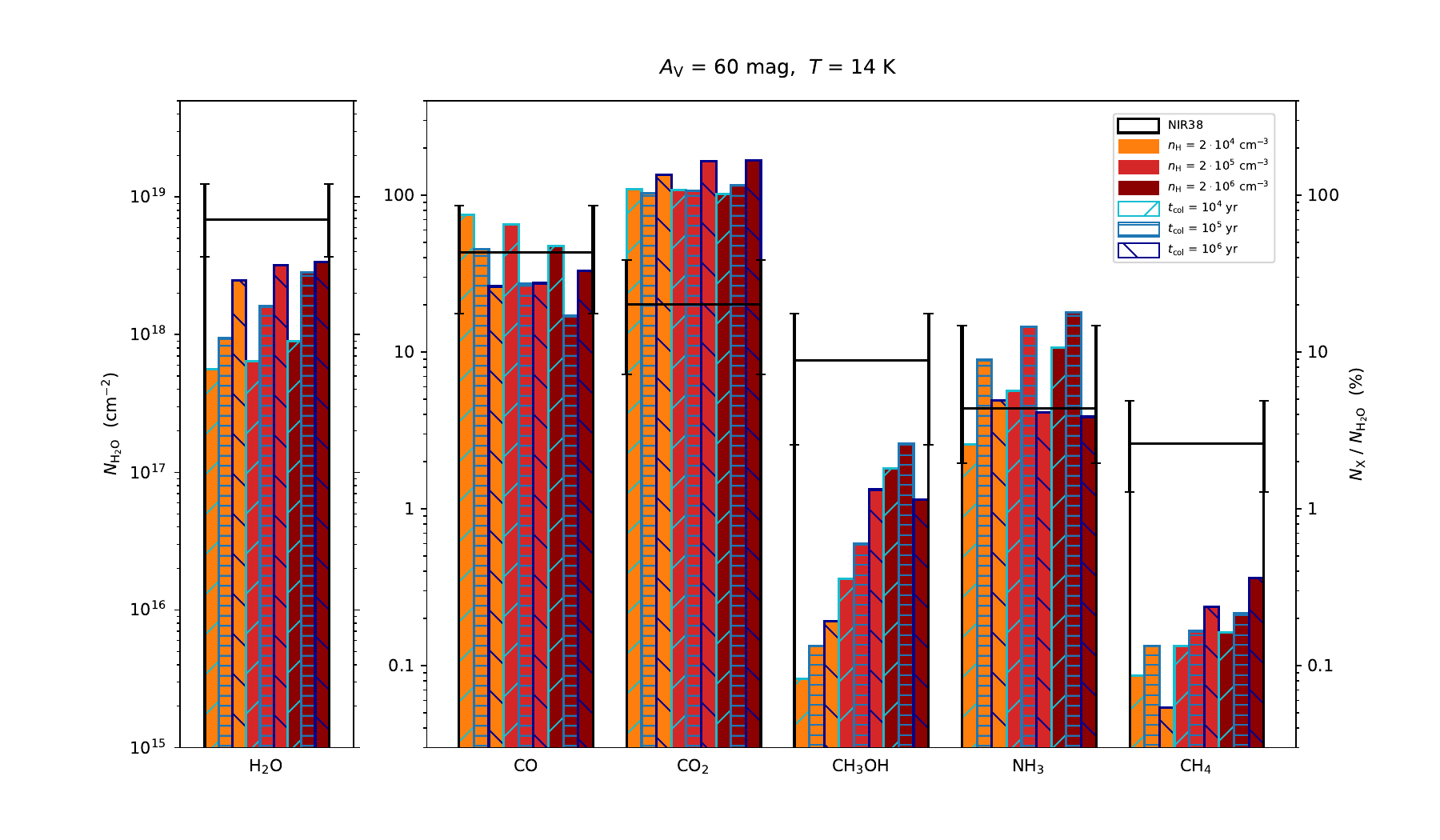}
\caption{Predicted ice compositions by \textit{MONACO} for NIR38. Details are as for Figure \ref{magickal-results-Av60}. \label{monaco-results-Av60}}
\end{figure*}

\begin{figure*}
\centering
\includegraphics[width=1\linewidth]{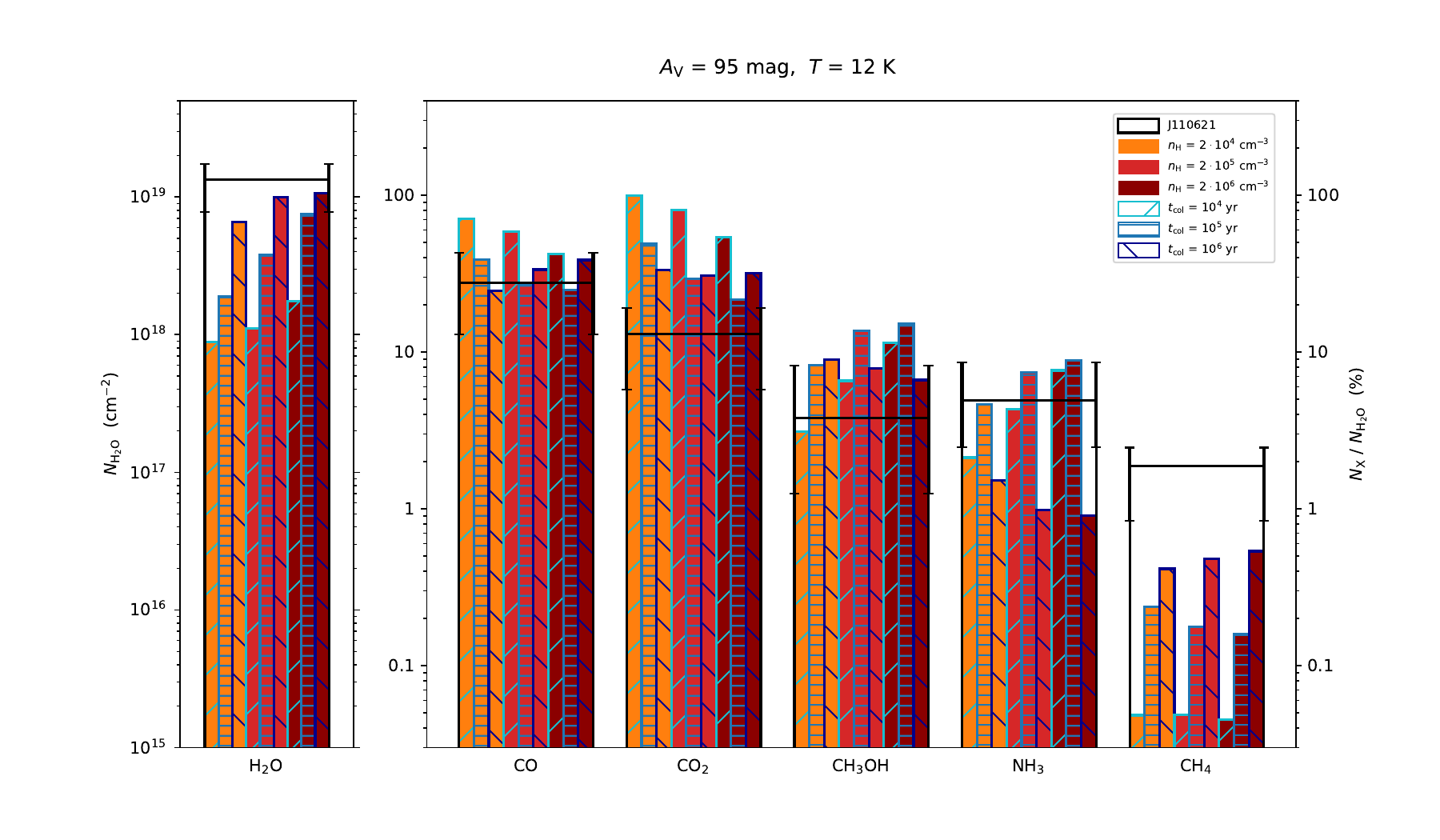}
\caption{Predicted ice compositions by \textit{MONACO} for J110621. Details are as for Figure \ref{magickal-results-Av60}. \label{monaco-results-Av95}}
\end{figure*}

\begin{figure*}
\centering
\includegraphics[width=1\linewidth]{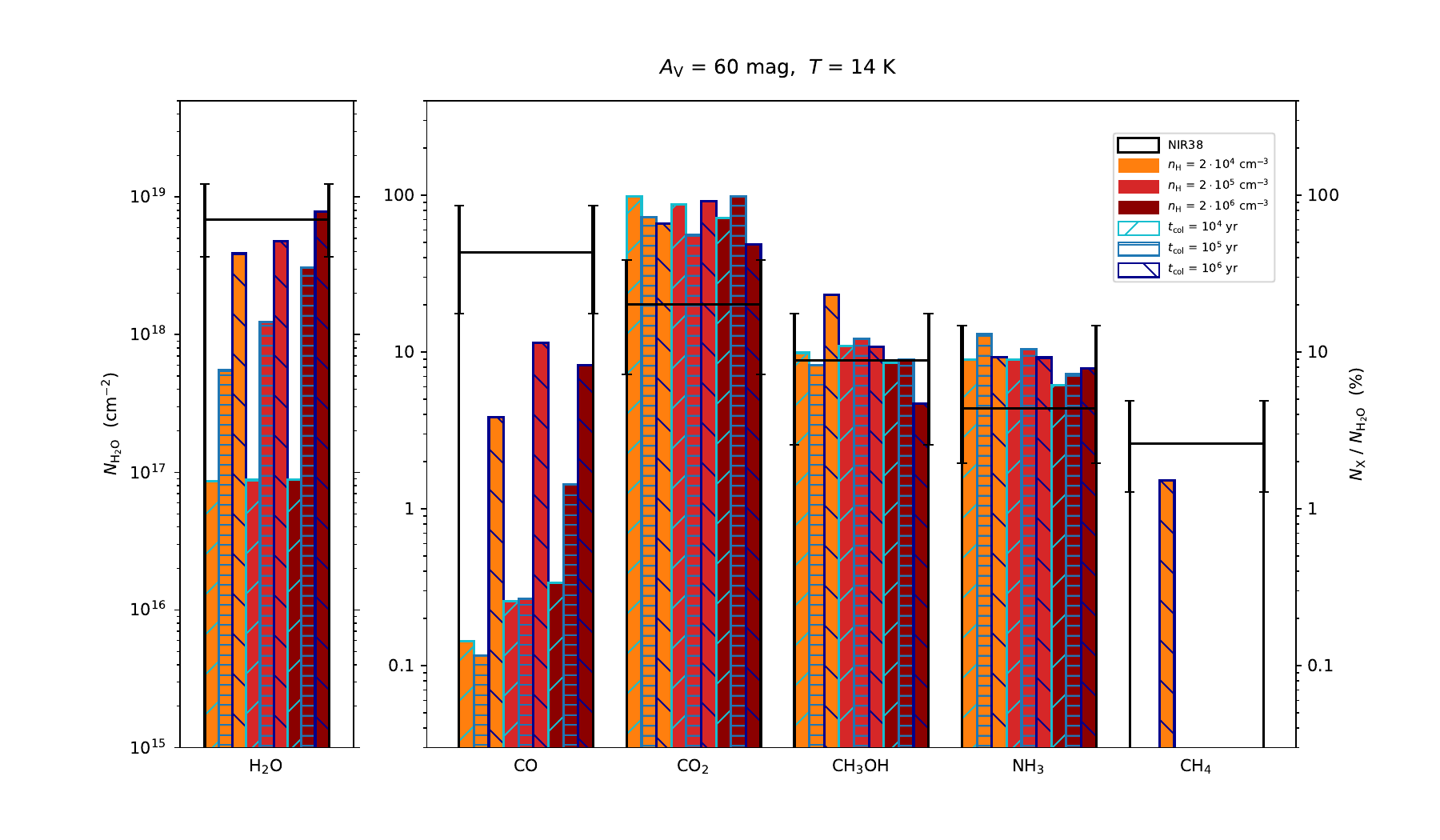}
\caption{Predicted ice compositions by Nautilus for NIR38. Details are as for Figure \ref{magickal-results-Av60}. Missing bars indicate ice abundance ratios $\leq$0.03\% relative to H$_2$O.} \label{nautilus-results-Av60}
\end{figure*}

\begin{figure*}
\centering
\includegraphics[width=1\linewidth]{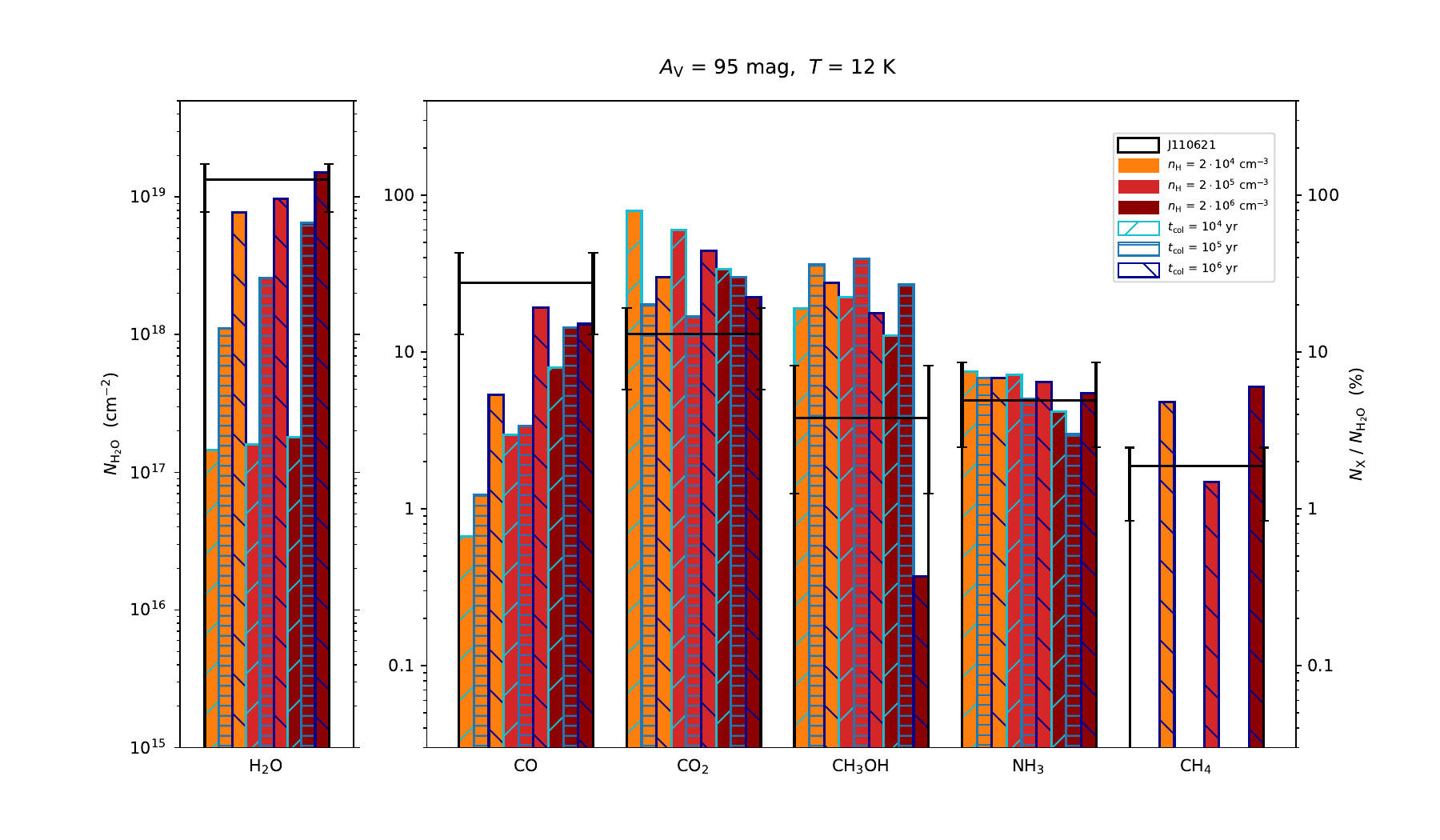}
\caption{Predicted ice compositions by Nautilus for J110621. Details are as for Figure \ref{magickal-results-Av60}. Missing bars indicate ice abundance ratios $\leq$0.03\% relative to H$_2$O.} \label{nautilus-results-Av95}
\end{figure*}

\begin{figure*}
\centering
\includegraphics[width=1\linewidth]{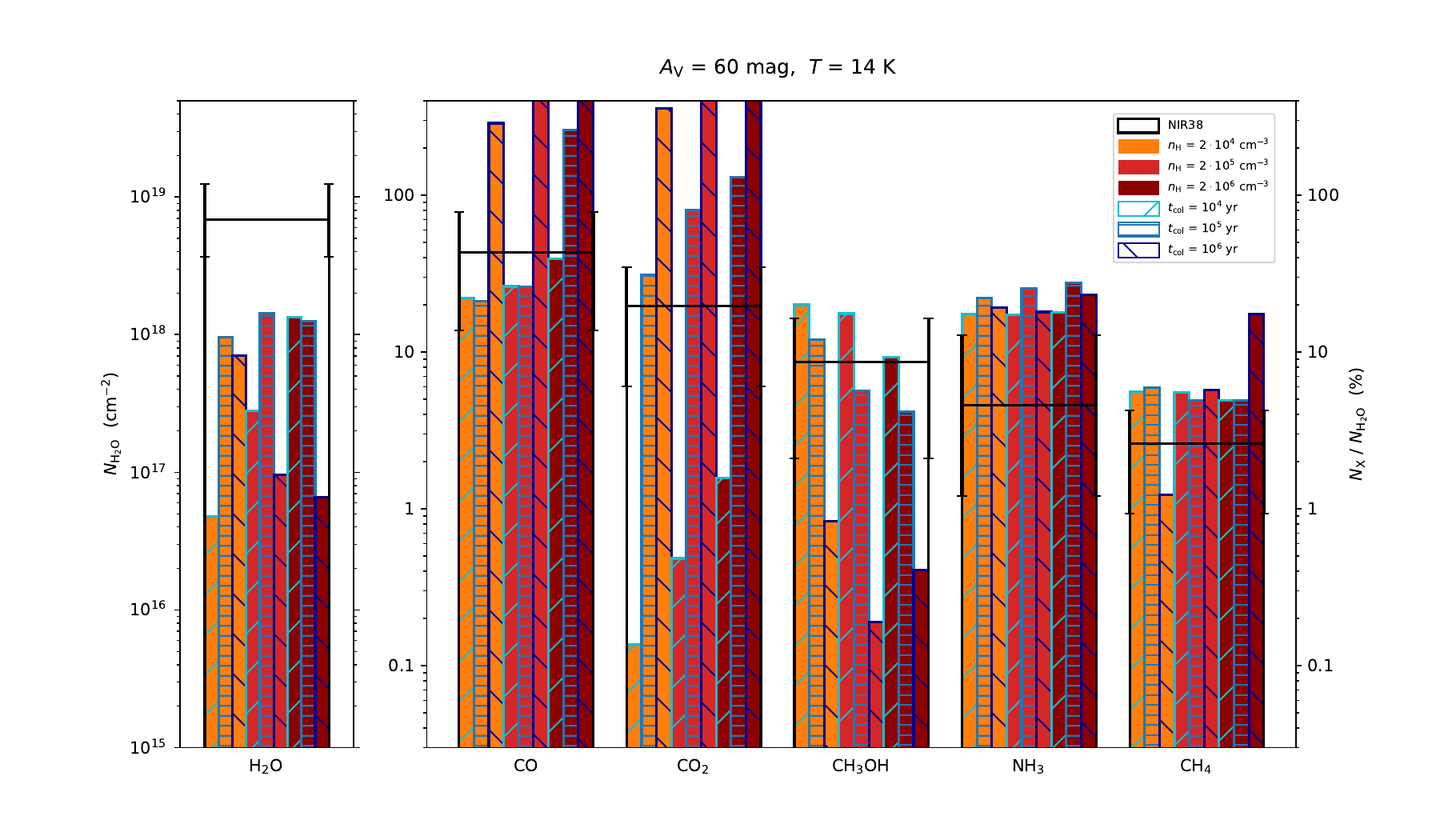}
\caption{Ice compositions predicted by \textsc{Uclchem} for NIR38. Details are as in Figure \ref{magickal-results-Av60}.}
\label{figure-uclchem-results-T14}
\end{figure*}

\begin{figure*}
\centering
\includegraphics[width=1\linewidth]{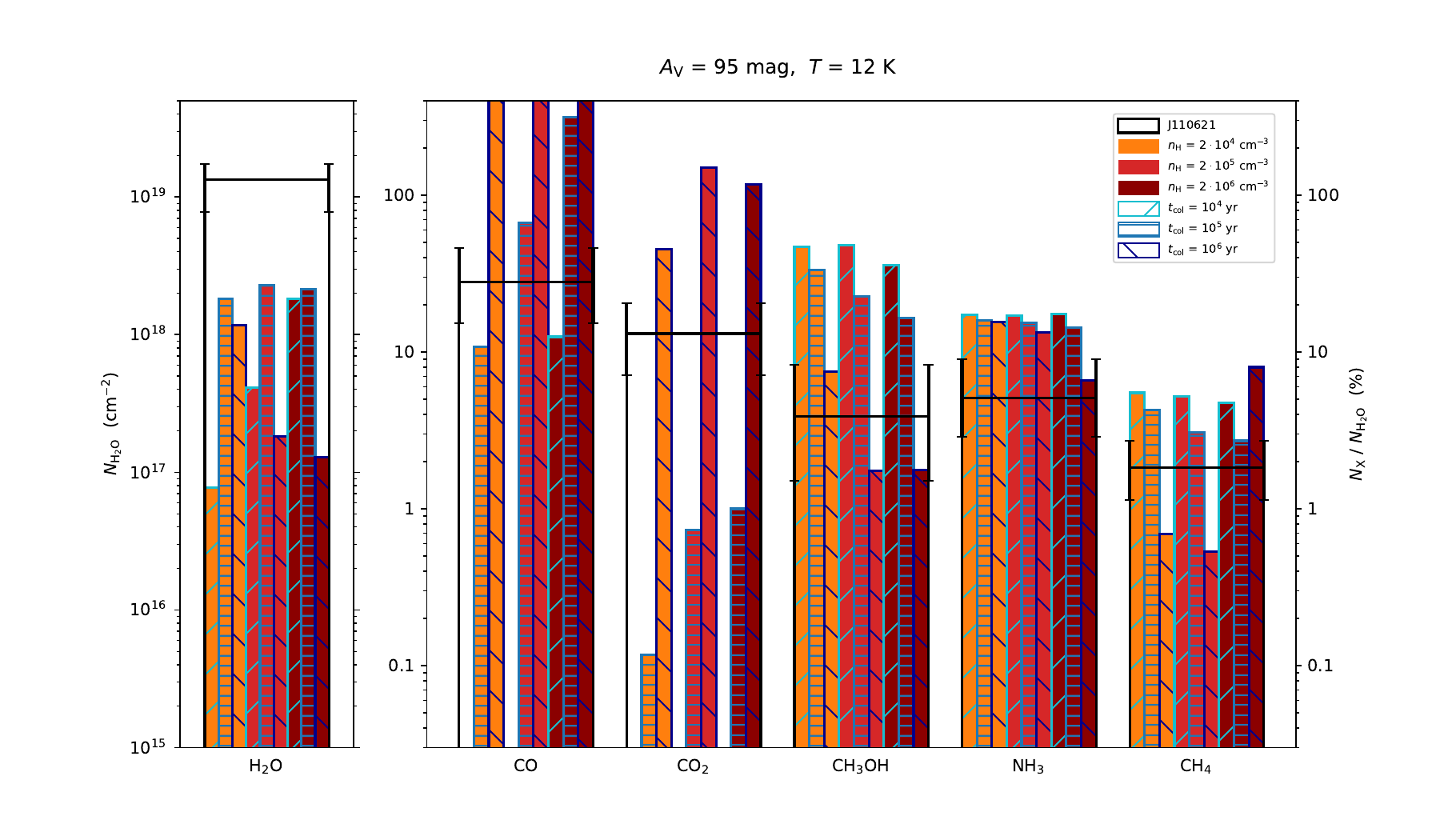}
\caption{Ice compositions predicted by \textsc{Uclchem} for J110621. Details are as in Figure \ref{magickal-results-Av60}. Missing bars indicate ice abundance ratios $\leq$0.03\% relative to H$_2$O.}
\label{figure-uclchem-results-T12}
\end{figure*}

\begin{figure*}
\centering
\includegraphics[width=1.0\linewidth]{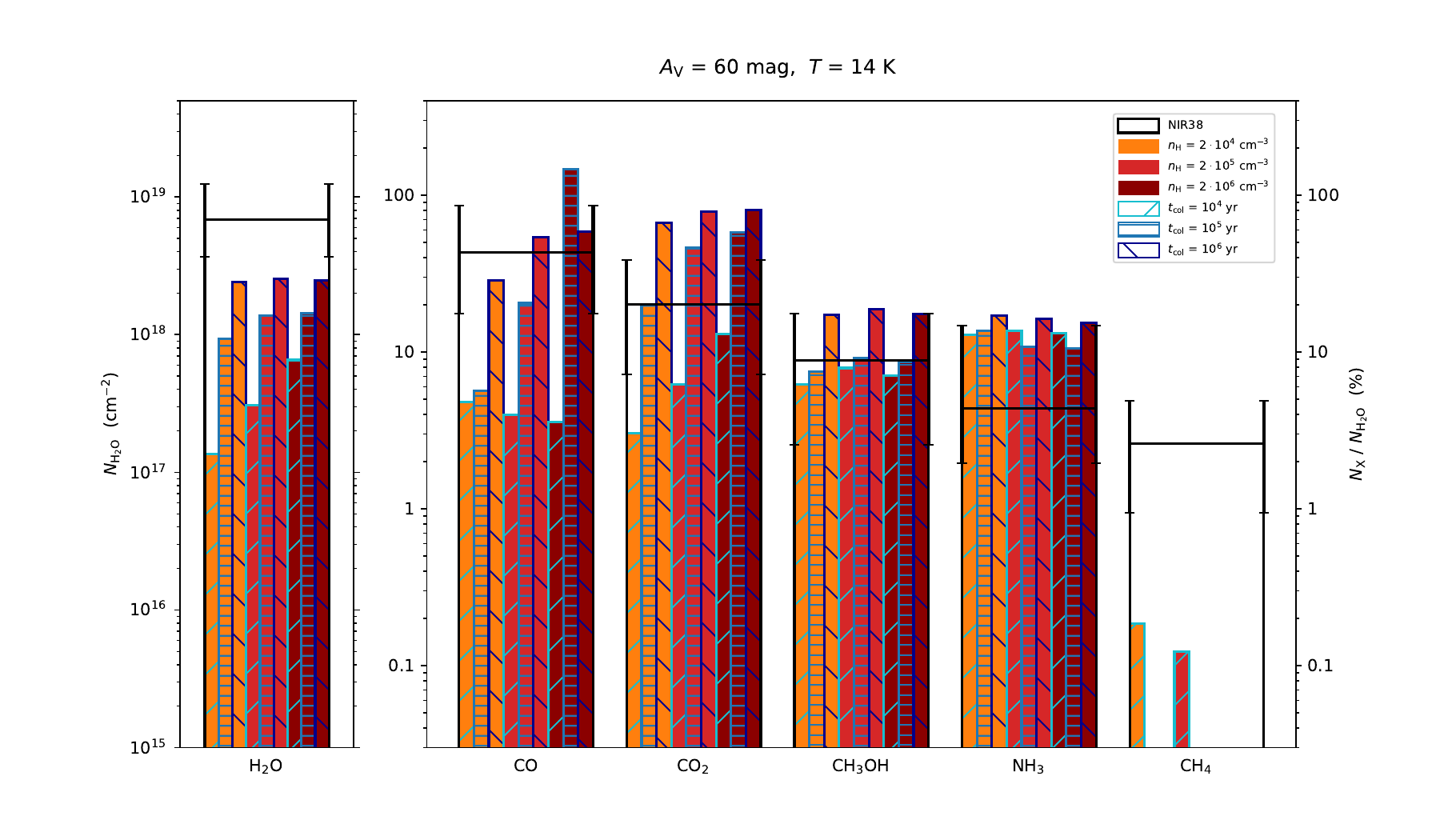}
\caption{Predicted ice compositions by the \textsc{KMC} simulations for NIR38. For models 6 ($n_{\rm H} = 2 \times 10^5\, {\rm cm}^{-3}$, $t_\textrm{col}$=10$^6$ yr), 8 ($n_{\rm H} = 2 \times 10^6\, {\rm cm}^{-3}$, $t_\textrm{col}$=10$^5$ yr) and 9 ($n_{\rm H} = 2 \times 10^6\, {\rm cm}^{-3}$, $t_\textrm{col}$=10$^6$ yr), the final abundances are taken from simulations that completed 96.8, 99.5 and 94.7 \% of their intended collapse times, respectively. Missing bars indicate ice abundance ratios $\leq$0.03\% relative to H$_2$O.} \label{figure-KMC-results-T14ALSO}
\end{figure*}

\begin{figure*}
\centering
\includegraphics[width=1.0\linewidth]{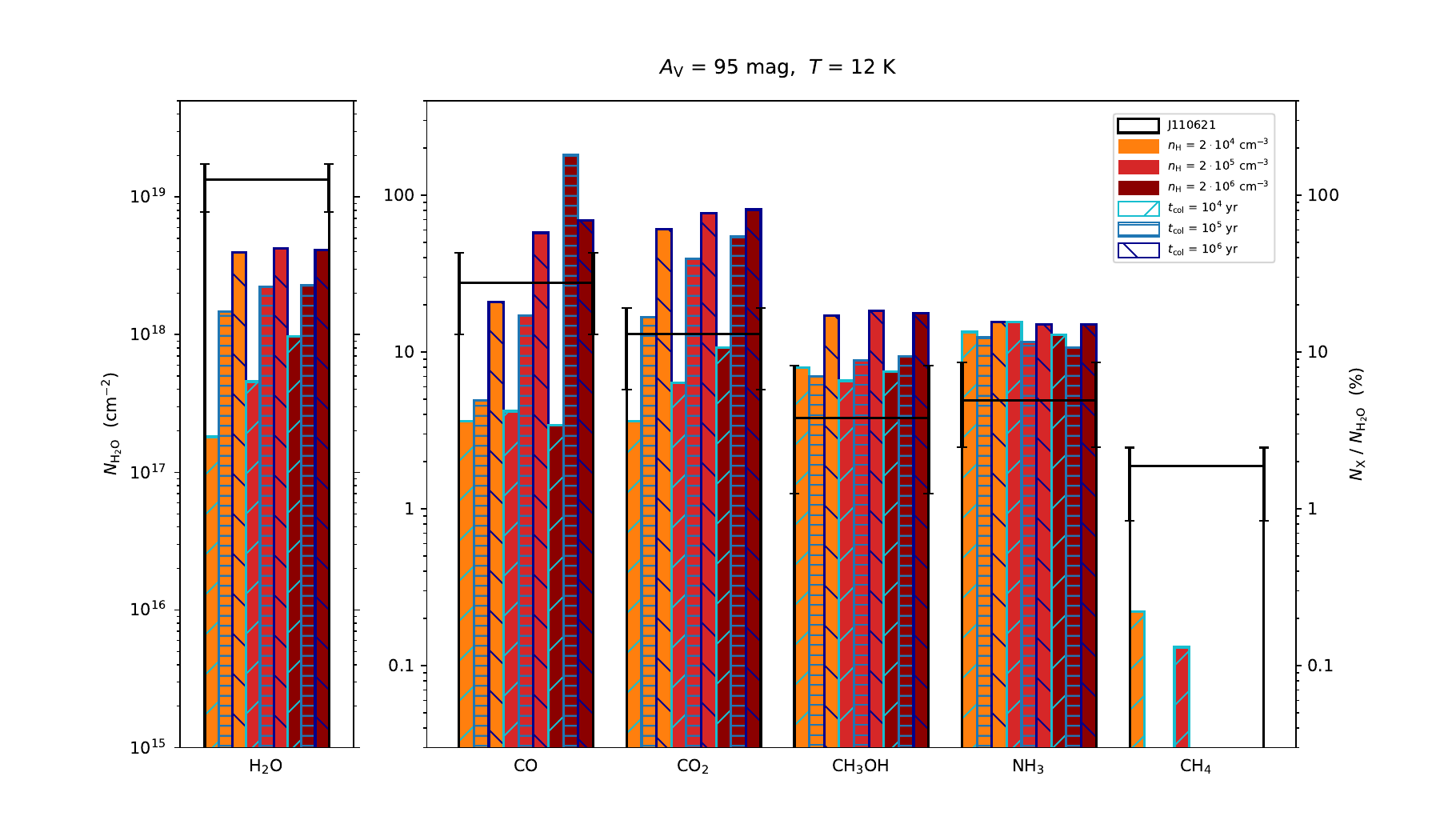}
\caption{Predicted ice compositions by the \textsc{KMC} simulations for J110621. Missing bars indicate ice abundance ratios $\leq$0.03\% relative to H$_2$O.} \label{figure-KMC-results-T12ALSO}
\end{figure*}

Figures$\,$\ref{monaco-results-Av60}-\ref{figure-KMC-results-T12ALSO} show a comparison between the ice composition predicted by the MONACO, Nautilus, \textsc{Uclchem} and KMC codes, and the ice abundances reported by \citet{McClure2023} toward the NIR38 and J110621 field stars.


\section{Evaluating the error of the models} \label{app:errors}

To evaluate the performance of the models in a systematic way, we used the same algorithm for all models. First, we select the models that predict \ch{H2O} column densities closer to the Ice Age observations, that is, the models with the lowest difference or error. Then, we calculate the mean absolute error (MAE) between the predicted and observed abundances with respect to \ch{H2O} on a logarithmic scale. Note that we have used the \ch{H2O} ice column densities for the first step of the MAE calculations, because the \ch{H2O} ice feature is the one with the highest signal-to-noise ratio in the JWST observations, and therefore, it presents the lowest associated uncertainties.

The equation used for the MAE is:
\begin{equation}
    \label{equation-mae}
    \mathrm{MAE} \;\;=\;\; \frac{1}{N} \; \sum_{i=1}^N \, \left | \, \mathrm{log}_{10} (y_i) \,-\, \mathrm{log}_{10} (\hat{y_i}) \, \right |   \;\; ,
\end{equation}
where $y_i$ represents the observed value of species $i$ (in abundance with respect to H$_2$O), $\hat{y_i}$ is the value predicted by the astrochemical model, and $N$ is the number of species observed with the JWST. We propagate the uncertainty in the values reported by \cite{McClure2023} to derive the uncertainty in the MAE. The best model corresponds to the one that presents the lowest MAE value within the estimated uncertainty coming from the observations.

\begin{figure*}
\begin{center}
\includegraphics[width=0.67\linewidth]{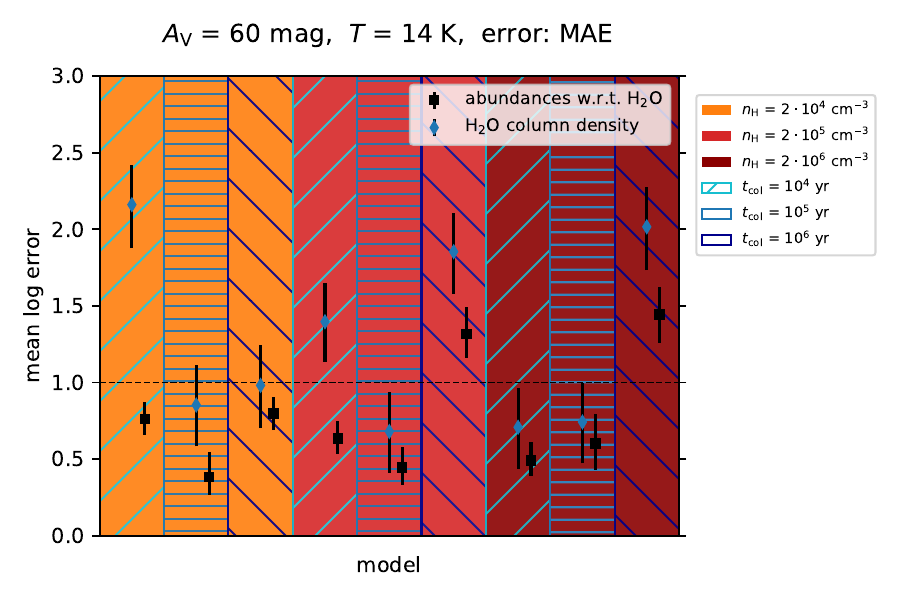}
\caption{Logarithmic errors between the predicted ice compositions by the different models of \textsc{Uclchem} for $A_\mathrm{V} = 60$ mag with $T_{\rm dust}$ = 14 K. Errors for only the \ce{H2O} column densities are indicated with blue rhombus, while black squares mark the errors calculated with the abundances with respect to water, both superimposed to the bars that represent the models in previous figures. Error bars correspond to 3$\sigma$.}
\label{figure-uclchem-errors-T14}
\end{center}
\end{figure*}

\begin{figure*}
\begin{center}
\includegraphics[width=0.67\linewidth]{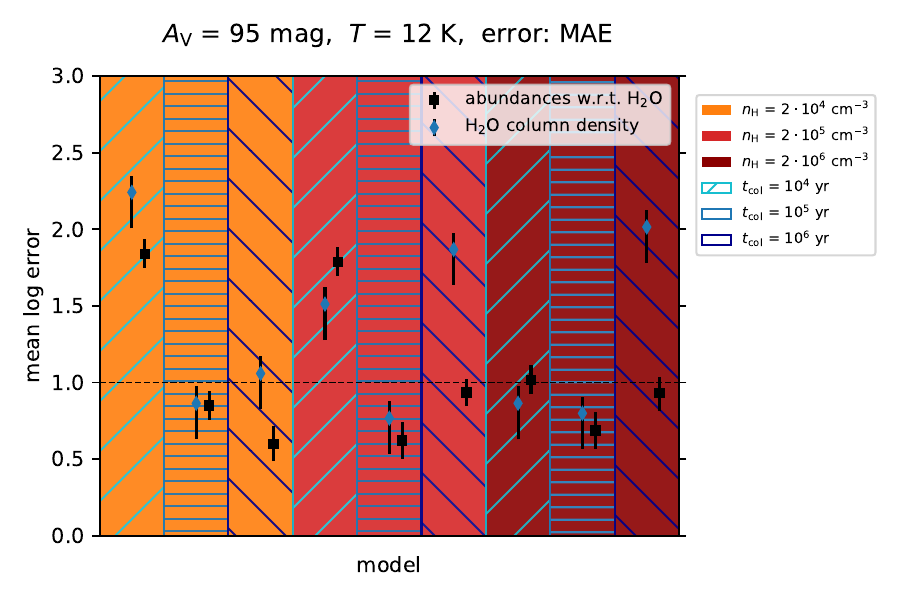}
\caption{Same as Fig. \ref{figure-uclchem-errors-T14}, but for J110621 ($A_{\rm V} = 95$ mag). The two best models correspond to intermediate collapse times and intermediate/high final densities (second and fifth models from the right).}
\label{figure-uclchem-errors-T12}
\end{center}
\end{figure*}

As an example, Fig. \ref{figure-uclchem-errors-T14} shows the \ce{H2O} error and the MAE for all the NIR38 models using \textsc{Uclchem}, while Fig. \ref{figure-uclchem-errors-T12} presents the same but for J110621. The error bars correspond to the 3$\sigma$ level. For NIR38, we can see that there are four models with the lowest MAE value (within the uncertainties), but the one with the lowest final density (2$\times$10$^4$ cm$^{-3}$) presents an error for \ch{H2O} column density which is too high and clearly larger than the rest, so we discard it. Then, the best models are the ones with intermediate collapse times ($t_{\rm col} = 10^5\, {\rm yr}$) and intermediate/high final densities ($n_{\rm H} = 2 \times 10^5,\,2 \times 10^6\, {\rm cm}^{-3}$), and also the one with high final density and fast collapse. However, if we now repeat the same procedure but with J110621 (Fig. \ref{figure-uclchem-errors-T12}), we obtain the same models except for the last one with fast collapse. Therefore, we conclude that the two best models for \textsc{Uclchem} are, for both sources, the ones with intermediate collapse times and intermediate/high final densities (second and fifth models from the right). 

We obtained analogous Figures for the remaining codes used in this paper. The results of the MAE analysis are reported in Table \ref{tab:mae} of Section \ref{sec:best}. We also note that we performed some tests using the root mean squared error (RMSE) and found similar results. The main difference is that the RMSE penalises the models that have bigger discrepancies between observations and predictions for some molecules. Therefore it would yield a big error if one species were very wrongly predicted, while it would favour models with a more uniform error between the different species predictions. For the RMSE, the equation used is:

\begin{equation}
    \label{equation-rmse}
    \mathrm{RMSE} \;\;=\;\; \left ( \frac{1}{N} \; \sum_{i=1}^N \, \left ( \, \mathrm{log}_{10} (y_i) \,-\, \mathrm{log}_{10} (\hat{y_i}) \, \right )^2 \right )^\frac{1}{2}  \; .
\end{equation}

\end{appendix}

\end{document}